\documentclass[twocolumn]{aastex63}
\usepackage{hyperref}
\usepackage{amssymb,amsmath}
\usepackage{refcount}
\newif\ifAMStwofonts
\def\sqiglt{\hbox{\rlap{\lower.55ex \hbox {$\sim$}}\kern-.05em \raise.4ex \hbox{$<$}\,}}
\def\sqiggt{\hbox{\rlap{\lower.55ex \hbox {$\sim$}}\kern-.05em \raise.4ex \hbox{$>$}\,}}
\def\til{\ensuremath{\sim\,}}

\def\chisq{\ensuremath{\chi^2}}
\def\rchisq{\ensuremath{\chi_{\nu}^{2}}}
\newcommand{\tim}[1]{\ensuremath{\times 10^{#1}}}
\def\deg{\ensuremath{^{\circ}}}
\def\etal{et al.\ }
\def\swift{\emph{Swift}}

\def\arcmin{\ensuremath{^{\prime}}}
\def\arcsec{\ensuremath{^{\prime\prime}}}
\def\cms{\ensuremath{$cm$^{-2}}}
\def\cstat{\ensuremath{\mathcal{C}}}
\def\xmm{\emph{XMM}}
\def\nh{\ensuremath{n_H}}
\def\lsrc{\ensuremath{L_{\rm src}}}
\def\lflat{\ensuremath{L_{\rm flat}}}
\def\ecf{erg \cms\ ct$^{-1}$}
\def\flux{erg \cms\ s$^{-1}$}

\begin{document}

\title{2SXPS: An improved and expanded Swift X-ray telescope point source catalog}
\shorttitle{The 2SXPS catalog}

\shortauthors{Evans et al.}

\author[0000-0002-8465-3353]{P. A. Evans}
\affiliation{University of Leicester, X-ray and Observational Astronomy Group, School of Physics and Astronomy,
University Road, Leicester, LE1 7RH, UK}
\email{pae9@leicester.ac.uk}

\author[0000-0001-5624-2613]{K.L. Page}
\affiliation{University of Leicester, X-ray and Observational Astronomy Group, School of Physics and Astronomy,
University Road, Leicester, LE1 7RH, UK}

\author{J.P. Osborne}
\affiliation{University of Leicester, X-ray and Observational Astronomy Group, School of Physics and Astronomy,
University Road, Leicester, LE1 7RH, UK}

\author{A.P. Beardmore}
\affiliation{University of Leicester, X-ray and Observational Astronomy Group, School of Physics and Astronomy,
University Road, Leicester, LE1 7RH, UK}

\author{R. Willingale}
\affiliation{University of Leicester, X-ray and Observational Astronomy Group, School of Physics and Astronomy,
University Road, Leicester, LE1 7RH, UK}

\author{D.N. Burrows}
\affiliation{Department of Astronomy and Astrophysics, Pennsylvania State University, University Park, Pennsylvania 16802, USA}

\author{J.A. Kennea}
\affiliation{Department of Astronomy and Astrophysics, Pennsylvania State University, University Park, Pennsylvania 16802, USA}

\author{M. Perri}
\affiliation{ASI Space Science Data Center, Via del Politecnico, I-00133 Rome, Italy} 
\affiliation{INAF-Osservatorio Astronomico di Roma, Via Frascati 33, I-000040 Monteporzio Catone, Italy}

\author{M. Capalbi}
\affiliation{INAF -- IASF Palermo, via Ugo La Malfa 153, I-90146, Palermo, Italy}

\author {G. Tagliaferri}
\affiliation{INAF-Osservatorio Astronomico di Brera, via E. Bianchi 46, 23807 Merate (LC), Italy }

\author {S.B. Cenko}
\affiliation{NASA/Goddard Space Flight Center, Greenbelt, MD 20771, USA}

\label{firstpage}

\begin{abstract} 
We present the 2SXPS (Swift-XRT Point Source) catalog, containing 206,335 point sources detected
by the \emph{Swift} X-ray Telescope (XRT) in the 0.3--10 keV energy range. This catalog represents
a significant improvement over 1SXPS, with double the sky coverage (now 3,790 deg$^2$), and
several significant developments in source detection and classification. In particular, we
present for the first time techniques to model the effect of stray light -- significantly reducing the number
of spurious sources detected. These techniques will be very important for future, large effective area
X-ray mission such as the forthcoming \emph{Athena} X-ray observatory. We also present a new model of the 
XRT point spread function, and a method for correctly localising and characterising piled up sources.
We provide light curves -- in four energy bands, two hardness ratios and two binning timescales -- 
for every source, and from these deduce that over 80,000 of the sources in 2SXPS are variable in at
least one band or hardness ratio. The catalog data can be queried or downloaded via a 
web interface at \url{https://www.swift.ac.uk/2SXPS}, via HEASARC, or in Vizier (IX/58).

\end{abstract}


\section{Introduction}
\label{sec:intro}

Serendipitous source catalogs have, for many years, been a standard product of
X-ray observatories giving great insights into the nature and range of X-ray
emitting objects in the Universe. Typically they can be divided into two
categories: large area but relatively shallow (such as the \emph{ROSAT} All-sky
Survey, RASS: \citealt{Voges99, Boller16}), or small area but deep (e.g.\ the
\emph{XMM-Newton} catalogs, \citealt{Watson09,Rosen16,Traulsen19}, and the
\emph{Chandra} catalogs, \citealt{iEvans10}). The output of the X-ray telescope
(XRT; \citealt{BurrowsXRT}) on the \swift\ satellite \citep{GehrelsSwift} lies
between these two extremes, and three point-source catalogs have been produced from XRT data.
SwiftFT \citep{puccetti11} focussed on the deepest ($>10$ ks) datasets, and 1SWXRT \citep{delia13}
analysed the individual observations; 1SXPS (\citealt{Evans14}, hereafter `Paper I') contained analysis
of both individual observations, and the combination of multiple, overlapping datasets.
1SXPS covered 1905 square degrees (nearly
double that of the more recent 3XMM-DR8 catalog), with a median 0.3--10 keV
source flux of 3.0\tim{-14} \flux, compared to 2.2\tim{-14} \flux\ (0.2--12 keV)
in 3XMM-DR8. Although XRT has a lower effective area (100 cm$^2$ at 1.5 keV) and
smaller field of view (radius 12.3\arcmin) than \emph{XMM}, it also has a much
lower background due to the orbital environment, which recovers much of the
comparitive sensitivity. Additionally, \swift\ observes a much larger number of
targets than is typical for a satellite, typically carrying out from tens to
hundreds of distinct pointings every day.

As well as providing a survey of moderate width and depth, the \swift-XRT data 
provide insight into the variability of the X-ray sky, since 95\%\ of its observations
are of areas of the sky which it has observed multiple times. Such information is critical
in the current era of time-domain astronomy, particularly multi-messenger astronomy,
to aid in the identification of X-ray counterparts to time-domain signals found
at other wavelengths or using non-photon triggers. For example, the localisations
of astrophysical neutrinos \citep{IceCube13} or gravitational waves \citep{Singer14}
are poor and many X-ray sources are found in follow-up observations \citep{Evans15,Evans16,Evans17}.
In order to correctly identify the true counterpart from the unrelated sources,
an understanding of the temporal properties of the serendipitous X-ray sky is crucial.

In this paper we present an updated \swift-XRT point source catalog: 2SXPS. 
This catalog contains 50\%\ more temporal coverage than 1SXPS, but contains 80\%\
more exposure (Table~\ref{tab:summary}), due to a change in which observations were selected for inclusion
(Section~\ref{sec:dataSelect}).

As well as updating the data in the catalog, we have updated our source detection system, 
focusing particularly on reducing the number of spurious detections due to diffuse emission
or stray light, as is discussed in some detail in Section~\ref{sec:bgsl}.

After we had begun processing the data for 2SXPS, \cite{Traulsen19} produced a catalog based on
stacking multiple co-located \xmm\ observations. As part of this work they demonstrated the use of
an adaptive smoothing technique combined with source masking as a means of estimating the
background, which they deemed more reliable in the presence of diffuse or structured emission than
the approach previously followed by the 2/3XMM and upon which the 1SXPS background modelling was
based. Due to the difference between the satellite orbits\footnote{\label{fn:ss}\swift\
is in a low-Earth orbit, so observations are comprised of one or more `snapshots'
(i.e. continuous exposures in a single orbit) of no more than 2.7 ks in
duration. Such snapshots are not perfectly aligned, meaning the background must
be modelled for each snapshot individually. The shortness of
the snapshots, combined with the smaller effective area of XRT compared to
\xmm-EPIC, and the different orbital environment of the two satellites results
in a much lower background in the individual XRT snapshots than in \xmm\
observations.}, background modelling
techniques from \emph{XMM} cannot be directly applied to \swift, but require a full simulation-based
investigation. Since the 2SXPS processing was in an advanced state when \cite{Traulsen19} was
published, such investigation is beyond the scope of this work and will be deferred to future anlaysis.

\section{Data selection, filtering and stacked image creation.}
\label{sec:dataSelect}

\begin{deluxetable*}{cccc}
%\begin{deluxetable}{ccc}
\tablecaption{Summary details of the catalog}
 \tablehead{
 \colhead{Category}                  &    \colhead{Value}               & \colhead{Units}          & Change from 1SXPS
}                                                                       
\startdata                                                              
Energy Bands:                        & Total:  $0.3\le E \le 10$        & keV         \\        
                                     & Soft: $0.3\le E < 1$ \\                               
                                     & Medium: $1\le E < 2$  \\                              
                                     & Hard:  $2\le E \le 10$ \\                               
Sky Coverage                         & 3,790                            & square degrees           & +99\%  \\
Time range                           & 2005 Jan 01 -- 2018 August 01    &                          & +52\%\\
Usable exposure                      & 266.5                            & Ms                       & +81\% \\
Number of observations               & 127,519                          &                          & +161\% \\
Number of stacked images             & 14,545                           &                          & +98\% \\
Median sensitivity$^1$ (0.3--10 keV) & 1.73\tim{-13}                    & \flux                    & -42\%\\
Median source flux (0.3--10 keV)     & 4.7\tim{-14}                      & \flux                    & +50\% \\
Number of detections                 & 1,091,058                        &                          & +86\%\\
Number of unique sources             & 206,335                          &                          & +36\% \\
Number of uncataloged sources$^2$    & 78,100                           &                          & +14\% \\                     
Number of variable sources$^3$       & 82,324                           &                          &+185\% \\ 
\enddata                                                                
\tablecomments{$^1$The flux at which source detection is 50\%\ complete at the median exposure
time. The 2SXPS source detection system is more sensitive than 1SXPS, however the median
exposure time in the catalog is also shorter which masks the true sensitivity gain. See Section~\ref{sec:sims} for more information.
The negative sign here shows that the 
2SXPS system has a \emph{lower} flux level i.e.\ improved sensitivity.
\newline 
$^2$Sources without a match within 3-$\sigma$ in any of the catalogs detailed
in Section~\ref{sec:xcorr} excluding the 2MASS, USNO-B1 and ALLWISE catalogs, as these have a high rate of
spurious matches. 
\newline 
$^3$Sources variable with 3-$\sigma$ confidence in at least one band or hardness ratio.
}
\label{tab:summary}
\end{deluxetable*}

We selected all\footnote{Excluding non-science observations with
target IDs beginning `0006'.} observations taken between 2005 January 01\footnote{Some of the data
taken prior to 2005, i.e.\ during spacecraft commissioning, have incorrect attitude information as
a result of commissioning work. These observations were included in 1SXPS by an oversight.} and
2018 August 01 with at least 100 s of exposure in the cleaned Photon Counting (PC) mode event list\footnote{i.e. those ending {\texttt \_cl.evt.gz} on the archive.}.
For the analysis in this catalog we used {\sc xrtdas}\footnote{\url{https://swift.gsfc.nasa.gov/analysis/xrt\_swguide\_v1\_2.pdf}}
v3.4.0 within {\sc heasoft} v6.22, and the most recent XRT calibration as of 2018 August 01.
All event lists were reprocessed using {\sc xrtpipeline} to give a self-consistent and up-to-date
dataset.

% Unlike 1SXPS we did not exclude
% observations where known large-scale diffuse emission occurs (e.g.\ Cas
% A), since the combination of the new \lflat\ statistic
% (Section~\ref{sec:Ltest}) and visual screening
% (Section~\ref{sec:screen}) were able to identify and often eliminate the
% false detections that such emission causes. 

The observations were
filtered to remove times where the data were contaminated by
scattered light from the daylight side of the Earth, and times when the
on-board astrometry was unreliable (determined by re-calculating the
astrometric solution using images from the UV/Optical telescope).
Details of this filtering were given in Paper I. Observations with
less than 100 s of PC mode data after such filtering were discarded from
the catalog.

Each selected observation was split into snapshots$^{\ref{fn:ss}}$; only snapshots of at least 50 s
exposure time\footnote{In 1SXPS we required at least 100 s.}  (after the above filtering) and at least one X-ray event
were included in the catalog. The pointing stability during the snapshot was also 
determined from the housekeeping data; if the pointing RA ($\alpha$) or declination ($\delta$)
had a standard deviation (from its mean) of more than 25\arcsec, the snapshot was discarded.
Any observation in which no snapshots passed these tests was excluded. This resulted in
127,519 observations in the catalog.

\begin{figure}
\begin{center}
\includegraphics[width=8.2cm]{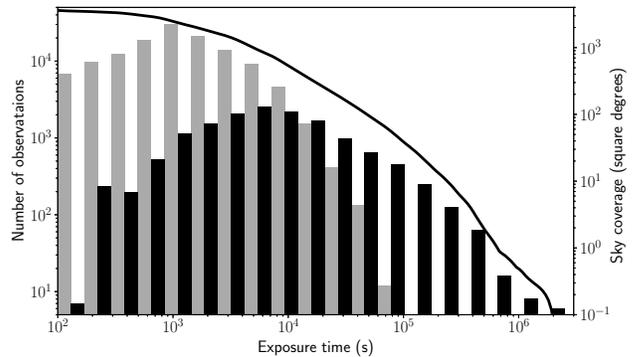}
\caption{The sky coverage and exposure details of the 2SXPS catalog. The solid line shows the sky coverage (corrected for overlaps)
as a cumulative function of exposure time (i.e.\ area with at least the exposure indicated). The histogram shows the distribution of exposure time
per dataset, with the individual observations shown in light gray and the stacked
images in black; the different colors are each half the width of the actual bins.}
\label{fig:area}
\end{center}
\end{figure}

As for 1SXPS, we also created `stacked images' in which all of the
observations of a given part of the sky were combined into a single
dataset for source detection purposes. This allowed us to maximise the
exposure time and hence sensitivity for each given point on the sky. In
1SXPS all images were limited to 1,000$\times$1,000 pixels
($\approx$ 40\arcmin$\times$40\arcmin). Since then, \cite{Evans15} developed tools to
allow XRT images to be stacked and analysed by our source-detection tools on an arbitarily-sized grid. For this
catalog we set the maximum stacked image size to be 2,300$\times$2,300
pixels ($\approx$ 90\arcmin$\times$90\arcmin) which corresponds roughly
to a 3$\times$3 grid of XRT pointings. This ensures that the processing
time of a given field remains managable, and that the co-ordinate
inaccuracy inherent in the tangent-plane projection co-ordinates used
for XRT data analysis is negligible. We created the minimum number of
stacked images necessary to ensure that every overlap between observations
is in at least one stacked image.
% We developed an algorithm to define
% stacked images such that the minimum number of images is produced
% necessary to ensure that, for each point of the sky observed by XRT, the
% maximum possible exposure is reached in at least one stacked image.
This yielded 14,628 stacked images. Throughout this work a
`stacked image' is as just defined, while an `observation' refers to the data
organised under a single obsid (which may comprise of multiple
snapshots, usually obtained within a single UT day).
The word `dataset' is used generically to refer to either an
observation or stacked image. 

The main characteristics of the 2SXPS catalog are given in Table~\ref{tab:summary}, along with a comparison with 1SXPS.
In Fig.~\ref{fig:area} we show the coverage of 2SXPS. The solid line shows the cumulative sky coverage as a function
of exposure time (corrected for overlaps). The histograms show the distribution of exposure time
in the individual datasets.

\section{Source detection}
\label{sec:sourcedet}

\begin{figure*}
\begin{center}
\includegraphics[width=16cm]{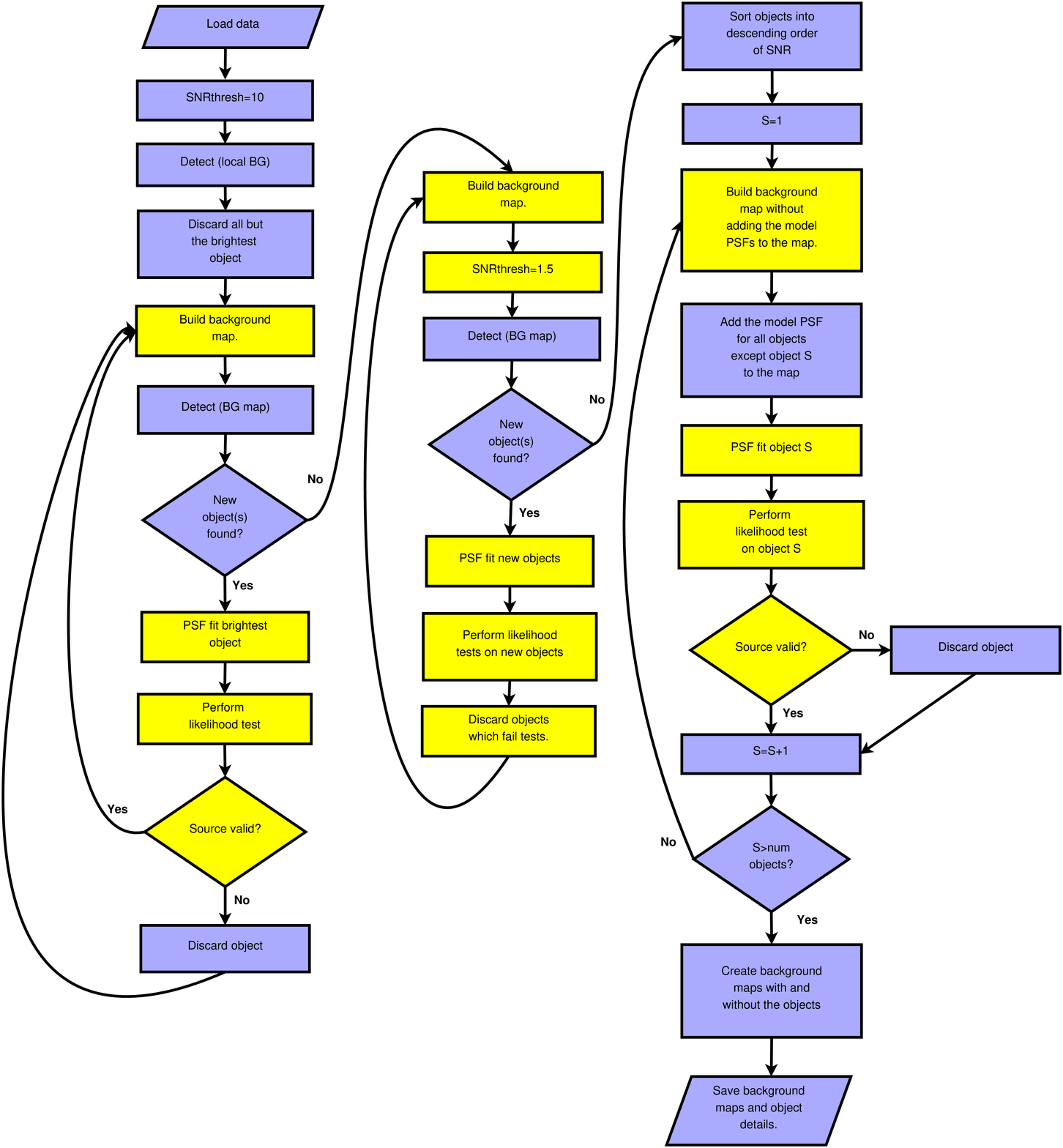}
\caption{Diagrammatic outline of the source detection process; the overall approach is as in 1SXPS
but important changes have been made in the boxes with yellow backgrounds; see text for details of these changes.}
\label{fig:algorithm}
\end{center}
\end{figure*}

The source detection system employed for 2SXPS was based on that described in Paper I with a
number of improvements. The algorithm for the detection phase is shown in
Fig.~\ref{fig:algorithm}; steps which are identical to their counterpart in 1SXPS have a lilac
background, whereas steps which were added or modified for 2SXPS have a yellow background. Here,
we briefly summarise the overall algorithm before discussing the modifications in more detail; for
an in-depth description of the overall approach see Paper I, section~3. We used the same algorithm
for both observations and stacked images, except where explicitly noted.

To prepare the data for source detection they were split into individual snapshots, for each of
which an exposure map and four FITS images were created -- one image per energy band in the
catalog\footnote{In Paper I, for stacked images of GRB fields we excluded the first snapshot --
when the GRB was likely to be very bright and piled up. Due to the improvements made for 2SXPS
(Sections~\ref{sec:psf}--\ref{sec:pup}) this was not necessary for 2SXPS.}. For stacked images,
the per-snapshot images and exposure maps were shifted onto a common sky coordinate frame
(Section~\ref{sec:coShift}). For each snapshot, the coordinates of the center of the XRT field of
view, the window size and the spacecraft roll angle were recorded\footnote{So that the background
map can be correctly constructed.}, and any potential sources of stray light were identified and
recorded (Section~\ref{sec:bgsl}). The per-snapshot images and exposure maps were summed to create
a single summed image per band and a single, summed exposure map. The source detection system was
then called, once per energy band; it made use of all of the files just described.

Unlike 1SXPS, the detection runs in the four bands were not entirely independent: information
about stray light and piled-up sources found in the total band (Sections~\ref{sec:bgsl}
and~\ref{sec:pup}) was passed to the other bands. However (as in 1SXPS), no other information was shared between
bands at this phase; this contrasts with the approach employed in the 2/3XMM catalogs
\citep{Watson09,Rosen16,Traulsen19}, where all bands were analysed simultaneously. This is because
XRT data must be split into snapshots to calculate the background map, which renders simultaneous
fitting across all bands computationally impractical. 
% Note that in a very small number of cases of
% short-exposure datasets one or more of the sub-bands contained no events, and so no source
% detection was carried out on that energy band.

The source detection process was a multi-pass process with three distinct phases, shown in the
three columns of Fig.~\ref{fig:algorithm}. It was based on a sliding-cell detection approach
combined with PSF fitting. At the start of phase one (left-hand column), an initial sliding-cell detection pass was
called for which the background was estimated from a box annulus around the sliding cell. This was
used purely to enable the creation of an inital background map (Section~\ref{sec:bgsl}).
Thereafter the remainder of phase one and all of phase two followed the same basic repeated
pattern: sliding-cell source detection, PSF fitting of the newly-detected source(s),
reconstruction of the background map with all detected sources first masked out and then the PSF
model of these added into the resultant map.

During the first phase, the signal-to-noise ratio (S/N) threshold for the sliding-cell detection,
defined in equation (6) of Paper I, was set to 10, and only a single source -- that with the highest
S/N -- was PSF fitted in each iteration. This reduced the number of spurious sources otherwise
found around bright sources. Once no S/N$>$10 sources could be found the second phase (middle column 
of Fig.~\ref{fig:algorithm}) began: the
S/N threshold was reduced to 1.5 and -- because these sources are less likely to yield spurious
sources in their wings -- all sources detected in each iteration were PSF fitted. In both of these
phases, likelihood tests were carried out on each PSF-fitted source (Section~\ref{sec:Ltest}), and
sources which did not achieve a status of at least \emph{Poor} were discarded.

Once no more sources were found in the cell-detect pass, the third phase (right-hand column 
of Fig.~\ref{fig:algorithm}) was carried out. Here, the PSF fitting was repeated for all
sources, using a background map containing the model PSFs of all
sources (except that being fitted), allowing a more accurate measurement
of each source's properties than was obtained in phases 1-2, where the source list
was incomplete and hence the background map inaccurate.

Once this process had been carried out on all datasets,
selected observations were manually inspected (Section~\ref{sec:screen}), and 
stray light issues were corrected with source detection repeated if appropriate.
Finally, the detections were combined into a unique source list (Section~\ref{sec:merge})
and then various source products were created (Section~\ref{sec:prods}).

Two statistics were used in various contexts throughout the fitting process: the C statistic
(\cstat, \citealt{Cash79}) as modified for use in {\sc xspec} \citep{Arnaud96} was the statistic minimised in fitting.
A so-called `likelihood'\footnote{The property referred
to as a `likelihood' in Paper I and the \xmm\ catalogs is not actually a likelihood,
or likelihood ratio in the normal statistical sense; it is just the negative of the natural log of a probability. Nonetheless we
retain the incorrect use of this term for ease, and consistency with previous work.}, $L$, was also calculated
at various stages to determine whether one fit was better than another.

\cstat\ was defined as: 

\begin{equation}
\cstat= 2 \sum_i \left(M_i - D_i + D_i\left[\ln D_i - \ln M_i\right]\right)
\label{eq:cstat}
\end{equation}

\noindent where $M_i$ is the model-predicted counts in pixel $i$, and $D_i$ is the actual number of 
counts in the pixel. 

The likelihood reflects the significance of an improvement in fit quality as a result
of adding in extra free parameters. Since $\Delta\cstat$ is distributed as $\Delta\chisq$,
the probability of the improvement arising by chance can be calculated, and the likelihood determined thus:

\begin{eqnarray}
L&=&-\ln P \nonumber \\
 &=&-\ln \left[\Gamma\left(\frac{\Delta\nu}{2},\frac{\Delta\cstat}{2}\right)\right] 
\label{eq:L}
\end{eqnarray}

\noindent where $\Delta\cstat$ and $\Delta\nu$ are the change in fit statistic and degrees of
freedom between the two fits respectively, and $\Gamma$ is the incomplete gamma function.

\subsection{Coordinate shifting for stacked images}
\label{sec:coShift}

For 1SXPS, only 4\% of the sky had been observed by overlapping
observations that, when stacked, produced an image larger than the
1000$\times$1000 pixel size limit in the standard software tools. Due to
new observing modes developed for \swift\ and used for
observational programs such as the follow up of neutrino detections
\citep{Evans15,Adrian-Martinez16}, the S-CUBED survey of the Small
Magellanic Cloud \citep{Kennea18}, the \swift\ Galactic Bulge Survey
\citep{Shaw17} and follow-up of gravitational wave events
\citep{Evans16c}; 42\%\ of the stacked
fields in this work were larger than this size limit. We therefore
created new software to shift the XRT images and exposure maps onto an
arbitrary co-ordinate grid. This software made use of the {\sc wcslib} C
library\footnote{\url{http://www.atnf.csiro.au/people/mcalabre/WCS/wcslib/}}
\citep{Greisen02,Calabretta02}; for each pixel in the original image,
{\sc wcslib} was used to convert the ($x,y$) co-ordinate into ($\alpha,
\delta$), and then again to re-convert this into ($x,y$) in the WCS
frame of the stacked image. For the data images, the integer ($x,y$)
positions of each event were converted into floating-point values and
randomly\footnote{Randomisation was performed using the {\sc gsl\_rng\_ranlxd1} 
random number generator provided by the GNU Scientific Libraries; the seed was
based on the computer clock time and process ID of the running task.}
positioned within their original pixel. For the exposure maps,
the four corners of the original pixel were translated as above to
identify the pixel(s) in the stacked image over which the exposure in the original
image should be distributed. This exposure was then shared among those pixels
according to the fractional
overlap. This method was based on the `area' transform method of the
{\sc swiftxform ftool}.

While this approach allowed arbitrarily-sized stacked images to be
created, there were limitations imposed by practical considerations, the
chief of which was computational efficiency. The computer resources
needed by our source-detection system scale approximately with the
number of snapshots, with an additional factor related to overall image
size. We therefore imposed a maximum image size of 2,300$\times$2,300
pixels ($\approx$ 90\arcmin$\times$90\arcmin) which is sufficient to
contain all observations within a standard \swift-XRT 7-point automated
mosaic, as commonly used for the follow-up of neutrino triggers, or
gamma-ray bursts detected by other satellites.

The data were split into stacked images based on their target IDs: a
unique, 8-digit identifier assigned to each target. In principle, all co-pointed
observations should have a common target ID, while all observations with
a common target ID should be co-pointed. The former constraint was not
always true, for operational reasons; however, this presented no difficulty, as co-pointed
target IDs were assigned to the same stacked image(s). The latter
constraint has occasionally been inadvertantly violated,
resulting in a small number of target IDs for which the different
observations have disparate pointings. For these cases, the observations
were split into co-pointed sets which were then assigned a new (unique)
target ID for the purposes of stacked image creation.

In order to ensure that the maximum sky depth was reached for each sky
location, target IDs could be assigned to multiple stacked images, and
stacked images could overlap. To demonstrate, consider the case of 4
adjacent target IDs along the same line of declination, spaced evenly so
as to slightly overlap each other; call these A, B, C, D. These would be
split into two stacked images, one comprising A, B and C; the other,
fields B, C and D. In this way all of the overlaps (AB, BC and CD) are
in at least one stacked image. The sky areas in targets B and C and the
overlap BC are in two stacked images, giving duplication of sources, but
duplication of sources is an inherent part of the catalog since
the observations making up targets A, B, C and D will all have also been analysed separately.
The rationalisation of the source lists is
described in Section~\ref{sec:merge}. In total, 2SXPS contains 34,553 targets
contributing to 14,545 stacked images; 7,260 of these target IDs contribute to more than one
stacked image. A further 4,022 targets exist which correspond to a unique observation
on the sky, and thus are in no stacked image.

\subsection{Background modelling and stray light}
\label{sec:bgsl}

During source detection, the background was repeatedly modelled
and the resultant `background map' was used by the sliding-cell
detection and the PSF fitting. The basic approach to background
modelling was identical to that in Paper I, which was based on that used
by the \xmm\ \citep{Watson09,Rosen16} and \emph{ROSAT}
\citep{Voges99,Boller16} catalogs. Sources already detected were
masked out, the data were coarsely rebinned and then this rebinned image
was interpolated back to each pixel in the image\footnote{For \xmm\ and
\emph{ROSAT} this last stage involved spline fitting, not interpolation.
The lower, less spatially-variable background of XRT is better handled
by interpolation.}. For any sources which had been PSF fitted in a
previous iteration, the PSF model was added to the background map,
reducing the likelihood of spurious sources being detected around a
bright source and enabling more accurate position determination of
nearby sources. This background modelling was conducted for each
snapshot separately, since the fields of view in each snapshot do not
exactly coalign. The resultant maps were then summed to give a single
background map for the dataset.

For 2SXPS we modified the approach from Paper I in two ways. First,
whereas in 1SXPS the background was rebinned into a 3$\times$3 grid, in
2SXPS for observations longer than 2 ks (i.e. with a better-sampled
background) a 5$\times$5 grid was used, enabling locally-elevated
backgrounds, for example due to diffuse emission, to be better modelled.
Second, stray light was included in the background map.

Stray light is an artifact of the Wolter-I optic design \citep{Wolter52}.
X-rays within the telescope's nominal field of view undergo two
grazing-incident reflections to focus them on the camera: off the
parabolic and then hyperbolic mirror surfaces. X-rays from
sources marginally outside of the field of view can also be scattered
onto the camera via only a single reflection off the hyperbolic
surface. Such X-rays fall in concentric rings on the detector (one ring
per mirror shell), referred to as `stray light'. This effect can be
predicted and analytically modelled, as described in detail in
Appendix~\ref{sec:app_sl}.

In Paper I stray light was handled manually, by eyeballing images, identifying regions affected
and flagging sources in those regions. For 2SXPS we developed a new technique to automatically
include stray light in the background map, dramatically reducing the number of spurious
detections. This consists of two main steps: first identifying sources capable of producing stray
light and the datasets in which stray light may be expected; then fitting the stray light in the affected images
and adding it to the background model.

\subsubsection{Sources of stray light}
\label{sec:sl}

\begin{figure}
\begin{center}
\includegraphics[height=8.1cm,angle=-90]{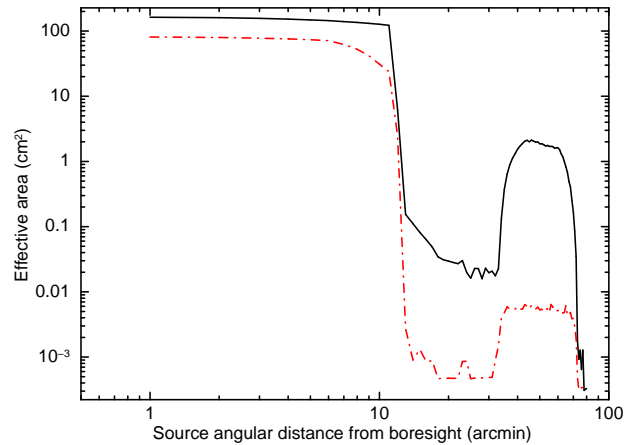}
\caption{The effective area of the XRT as a function of off-axis angle, derived
from ray-tracing simulations. The focussed field of view is 12.3\arcmin\ in radius. The
solid black curve shows the total effective area for a point source, integrated
over the entire CCD, as a function of off-axis angle. The broken
red line shows the peak effective area, which occurs in a CCD area equivalent to the
half-energy-width of the PSF (18\arcsec\ diameter). The two curves are more
disparate for sources outside the nominal field of view, because the X-rays for these are spread out over
a much larger area of the CCD in large rings, whereas for on-axis sources the
counts are focussed into the spot-like point spread function.}
\label{fig:slarea}
\end{center}
\end{figure}

The predicted effective area of the XRT as a function of off-axis angle, derived from ray tracing,
is shown in Fig.~\ref{fig:slarea}. This agrees with the in-flight measurements of 
\cite{Moretti09}. Any source \til35--75\arcmin\ off
axis will produce stray light in the XRT; however, for most sources this
will be so weak and diffuse as to be irrelevant. In 1SXPS the median
0.3--10 keV background rate was 8.6\tim{-7} ct s$^{-1}$ pixel$^{-1}$;
thus, only sources bright enough to produce stray light at around this
intensity need be considered. The half-energy-width (HEW) of the XRT PSF
is 18\arcsec; the red dashed line in Fig.~\ref{fig:slarea} corresponds to this, i.e. a
region 45 pixels in area, in which the mean 1SXPS background level would
be 3.9\tim{-5} ct s$^{-1}$. The ratio of on- and off-axis effective
areas in such a region width is \til3\tim{-5}, and by definition the
true source count rate of an on-axis source is double the count rate
measured in an HEW region. So a source with an on-axis
count rate of 2.7 ct s$^{-1}$ or higher can, when 35--75\arcmin\ off axis,
contribute photons to the XRT at a level similar to the normal background.

We compiled a list of all sources in 1SXPS with a count rate above this
level. Using {\sc pimms} and assuming a typical AGN spectrum (a
power-law with $\Gamma=1.7$, \nh=3\tim{20}\cms) to convert this into
expected brightnesses in \xmm\ and \emph{ROSAT}, we added to this list
any source in the HEASARC X-ray Master
catalog\footnote{\url{https://heasarc.gsfc.nasa.gov/W3Browse/all/xray.html},
queried on 2017 July 1.} above this flux. We also added all the contents of the
\emph{INTEGRAL} reference catalog \citep{Ebisawa03}, queried via
HEASARC on 2017 July 1: this catalog contains any sources ever
recorded above \til1 mCrab ($\approx$ 1 ct s$^{-1}$ in XRT) at 3keV.
This list was then consolidated to remove duplicates and provided a
reference list of possible stray-light sources. When the data were split into snapshots,
any source in this list which lay
31\arcmin--72.5\arcmin\ away from the center of the XRT field of view was recorded
as a possible source of stray light. Because the field of view can vary by several arc~minutes between
snapshots, this check was done for each snapshot independently.

\subsubsection{Including stray light in the background maps}
\label{sec:slmodel}

In principle, if we know the position of a source with respect to the
XRT boresight and its intrinsic flux, the expected stray light from the
source can be calculated analytically, as described in
Appendix~\ref{sec:app_sl}, and then added to the background map. In
practice this cannot be done for two reasons: first, only the 1SXPS and
2/3XMM sources have positions accurate enough for this to be done
`blind'; second, many of the sources are variable and their intensity at
the time of the XRT observations is not known. Additionally, the
analytical model is not perfect and sometimes the data were better
modelled using a slightly incorrect source position. We therefore fitted
the predicted stray light to the image.

We defined three free parameters per stray light source: $\theta, \phi,
N$. The first two represent the source position (as position angle
relative to the CCD DETX axis, and angular distance off-axis  respectively), the third was
its normalization. This fitting is a somewhat involved process due to
three chief complications. First, unmasked point sources in the image
can dominate the fit, resulting in very poor reproduction of the stray light. Conversely,
since stray light gives rise to spurious detections, masking out point
sources can result in the stray light being entirely masked out and so
unfittable. Second, the stray light contribution should be separated
from the underlying background, otherwise the rebin/interpolate approach
to creating the background map will overestimate the background in
regions near to the stray light. Third, the pointing direction can vary
by several arc minutes between snapshots, which is sufficient to
significantly change the stray light pattern. Due to the inaccuracies in
the model (Appendix~\ref{sec:app_sl}), it is not sufficient to identify
the position and normalization of the stray-light-causing source in one
snapshot and then simply adjust the parameters according to the pointing
differences.

The algorithm developed to surmount these issues and provide a model
of the stray light is described in Appendix~\ref{sec:app_slfit}. It was optimised by running it on a
series of 1SXPS datasets with and without stray light. Even so, a visual check was made
of all possible stray-light fields, as described in Section~\ref{sec:screen}.

\subsection{PSF model and fitting}
\label{sec:psf}

The PSF fitting of new sources proceeded largely as in Paper I: a
circular region was identified, centered on the cell-detect position and
with a radius depending on the source S/N, the source position ($x,y$)
and normalization were identified as free parameters, and \cstat\
(Equation~\ref{eq:cstat}) was minimised. Minimization was carried out
using the {\sc MnMinimize} class in the {\sc minuit2} C++
libraries\footnote{\url{http://project-mathlibs.web.cern.ch/project-mathlibs/sw/Minuit2/html/index.html}}.
The position uncertainty was found for each axis independently, by
stepping the position in that axis and refitting (while keeping the test
position frozen); where \cstat\ increased by one from the best-fitting
value gave the 1-$\sigma$ confidence interval on the position in that
axis. The radial position error was determined from this,
by taking the mean of the RA and Dec errors, and then converting this to a 90\%\
error using Rayleigh statistics\footnote{\label{fn:errerr} In the initial
catalog reased on 2019 November 26, an error had been made in the conversion to 90\%\ resulting in the \emph{statistical}
errores being a factor of 1.5 too high. This had no material impact on the catalog contents,
and was fixed on 2020 Feb 11.}. 
Very occasionally in 2SXPS, the position error could not be found in this
way: {\sc MnMinimize} failed to return a valid fit while the source
position was being stepped around. In this case the radial position
error of the source was set to $\frac{11.301\arcsec}{\sqrt{N}}$ (90\%\
confidence), where $N$ is the number of counts in the PSF fit; this
relationship giving the best-fit to the 2SXPS position errors determined
succesfully by \cstat\ stepping. There were also some cases where
position errors were found, but were much smaller for the number of
counts than was typical of the catalog. Such values may indicate that
the \cstat\ stepping encountered difficulties, but equally there are
cases (e.g.\ crowded fields) where \cstat\ can vary sharply with
position. We did not alter these small values.

A few changes from the Paper I
approach need to be noted. First, sources with a S/N $\ge$60 from the
cell-detect phase were fitted over a region with a radius of 40 pixels (in Paper I
everything with S/N$\ge$ 40 had a radius of 30 pixels). Second, if the
position returned by the PSF fit had moved from the input position by
more than 50\%\ of this radius, the fit was repeated using a
new region centered on the new position. A source could be refitted in this way
no more than 5 times (to prevent infinite loops if a degenerate position
solution was found). This was beneficial because, for very piled up
sources where the PSF core has no counts in it (see
Section~\ref{sec:pup}), the true source position could lie outside of
the initial PSF-fit region. A third change relates to the way pile up
was handled, and will be discussed in the next section.

As well as these procedural changes, we considered the shape of the XRT
PSF. Within the \swift\ software and calibration database (CALDB) the PSF is defined
as the combination of a Gaussian and King function:

\begin{eqnarray}
P(r)&=&N e^{\frac{r^2}{2\sigma^2}} + \left(1-N\right) \left[1+\left(\frac{r}{r_c}\right)^2\right]^{-\beta}
\label{eq:psf}
\end{eqnarray}

\noindent where $N$ is a normalization, $r$ the radius at which the PSF is evaluated, 
and $\sigma, r_c$ and $\beta$ the parameters controlling the shape of the Gaussian and King components.
\cite{Moretti05} calibrated this
in flight and deemed that only the King-function component (the second
part) was necessary, i.e. $N=0$. While this proves a good
description of most sources, we have found that for bright objects the
outlying wings of the PSF appear to be underpredicted by this model, consistent
with fig.~5 of \cite{Moretti05}. This results in the background map around bright sources being too low
and spurious sources being detected around bright objects. In Paper I we
handled this by defining a `blind spot' around bright sources, in which
detections were discarded as likely duplicates of the central object.
This approach is less than ideal, as real objects do appear near
bright ones. For this paper, we therefore attempted a recalibration of
the PSF, in order to better model the wings. This work is described in
Appendix~\ref{sec:app_psf}, and produced the PSF parameters shown in
Table~\ref{tab:psf}. This PSF was used throughout this work, and will replace the existing
PSF definition in a future {\sc caldb} release. This
dramatically reduced the number of spurious detections around bright sources.
In Paper I (Appendix A), we derived a function to model the `spokes' in the PSF (the shadows of the mirror support structure):
this is a function of PSF radius and azimuthal angle and the original, azimuthally symmetric PSF model is multiplied by this function. 
This function is not affected by the new PSF definition and was used as in Paper I.

\begin{deluxetable}{cc}
\label{tab:psf}
\tablecaption{The PSF parameters derived for and used in 2SXPS. The PSF is defined in
Equation~\ref{eq:psf}.}
\tablehead{
\colhead{Parameter}      &  \colhead{Value}   
} 
\startdata
N             & 0.080 \\
$\sigma$      & 3.119 pixels (=7.351\arcsec) \\
$r_c$         & 1.597 pixels (=3.764\arcsec)\\
$\beta$       & 1.282 \\
\enddata
\end{deluxetable}

\subsection{Pile up}
\label{sec:pup}

\begin{figure*}
\begin{center}
\includegraphics[width=15cm]{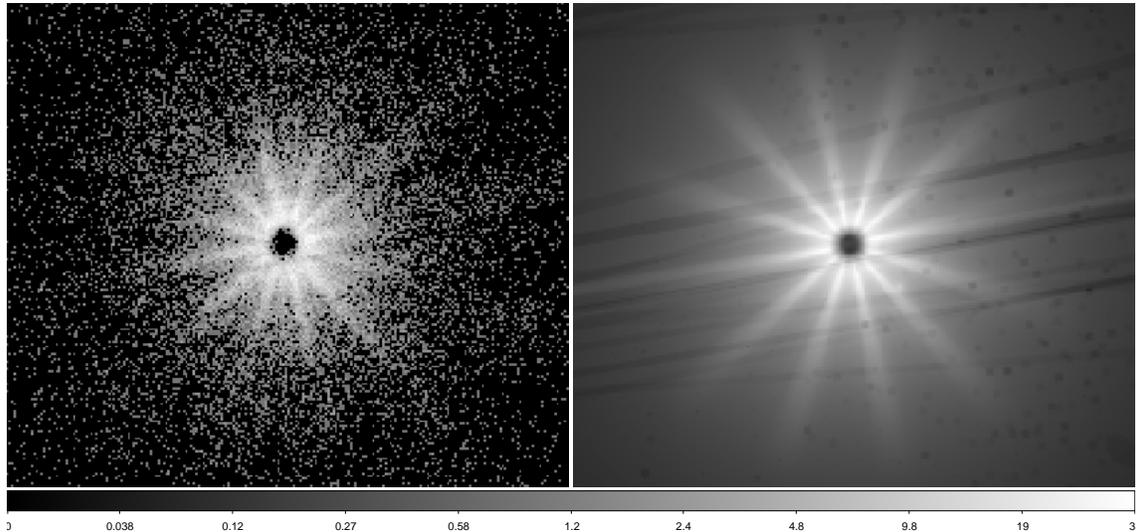}
\caption{The effect and successful modelling of pile up. \emph{Left}: An extremely piled up source,
4U 1820$-$30, in which the centre of the PSF contains no counts due to grade migration (see text).
\emph{Right}: The background map of this dataset containing the fitted PSF model of the source, showing
that pile up has been well reproduced.}
\label{fig:pupfit}
\end{center}
\end{figure*}

Pile up is a phenomenon affecting photon-counting detectors such as the
XRT. It occurs when multiple photons impact the same or adjacent CCD pixels within a single
exposure frame, and on read-out the charge thus liberated is interpreted as arising from a single
photon. Since this is a stochastic process, some fraction of the events from any source
will be affected by pile up; however, this fraction only becomes significant at moderate
source intensities: in XRT PC mode, pile up typically starts to become a factor
for sources around 0.6 ct s$^{-1}$.
Pile up is initially manifested by the core of the PSF being slightly suppressed compared to the wings, and the
source spectrum being artificially hardened. A second factor is
so-called \emph{grade migration}: events are assigned a grade based on
how many adjacent pixels are affected by the cloud of charge liberated
by the incident X-ray. In the case of pile up, separate X-rays incident
on adjacent pixels will be erroneously recorded as a single event
covering both pixels. Once pile up becomes severe, this causes events to
have invalid grades\footnote{That is, grades above 12 for PC mode, see
\url{https://www.swift.ac.uk/analysis/files/xrt_swguide_v1_2.pdf}.} and
thus be rejected, resulting in an apparent `hole' in the core of the PSF; an example
of such a source is shown in Fig.~\ref{fig:pupfit}.

\cite{Evans09} developed a series of discrete PSF profiles whereby a Gaussian component was
subtracted from a King component, which approximately described the PSF at increasingly degrees of
pile up. In Paper I, each of these profiles in turn was applied to a source, and the most
appropriate profile was determined based on the fit statistic. Since we have redefined the PSF for
this work (Section~\ref{sec:psf}), these old profiles can no longer be used; and because the new
PSF has both King and Gaussian components, the addition of a third element would also be
incompatible with the existing CALDB and {\sc xrtdas} software. However, pile up can be very
accurately modelled simply by multiplying the PSF by an analytical multiplicative loss function.
This function was original used by \cite{Popp00} to describe the spectral energy redistribution of the \xmm\ EPIC-pn
camera but works well in our context. It depends only on radius $r$ and is given by:

\begin{eqnarray}
f(r) =  S + B \left( \frac{r}{l} \right)^ c  :: (r < l) \\
f(r) =  1 - A e^{-\frac{r-l}{\tau}}  :: ( r \ge l)  
\label{eq:pup}
\end{eqnarray}

\noindent where 

\begin{eqnarray}
\label{eq:pup2}
B = \frac{ l (1-S) }{l + c \tau} \\
A = 1 - S - B
\end{eqnarray}

\noindent provided $S<1$; otherwise $A=B=0$ and the function has no
effect. Thus, pile up can be modelled by the addition of the following
four free parameters. $S$, which can be in the range $[0,1)$ determines
the overall depth of the loss function: for $S=0$ there is a hole in the
center of the PSF, at $S=1$ pile up has no effect. $l$, which was
limited to $[0.1,50]$, controls the overall scale of the loss function,
and the transition from the core to the wings of the PSF. $c$, which we
restricted to $[0.1,10]$, affects the steepness with which the
loss function changes in the PSF core, and is complemented by $\tau$,
which could cover $[0.1,200]$ and controls the loss function out in the
PSF wings. With the exception of $S$ we had no \emph{a priori}
expectations of what ranges the parameters should cover, and the above ranges
were taken as those which (from CCD simulation work, Beardmore \etal\ in prep) could be
deemed reasonable.

When performing a PSF fit to a source, the fit was originally carried out using
the new PSF model (Equation~\ref{eq:psf}) with no loss function. If the source
had a S/N from the cell-detect pass of at least five, a second PSF fit
was performed, this time with the loss function included, and hence four
extra free parameters. As with the original fit, if this moved the
position significantly, the fit was repeated with a new region centered
on this position: see Section~\ref{sec:psf}. The likelihood value
relating to the new fit was calculated using Equation~\ref{eq:L}, where
$\Delta\cstat$ was the difference between the with/out loss-function
fits, and $\Delta\nu=4$ (the loss function parameters). If $L>10$ the
source was deemed to be piled up, and the results of the fit with a
loss function were taken as the source parameters.

If a source is affected by pile up, the PSF shape will be affected in
all energy bands (although not necessarily to the same degree, as pile up
causes soft events to migrate to the harder energy bands), regardless of
the brightness in that band, which can cause problems for the algorithm
as described above. Consider for example, a very piled-up, very absorbed
source. There may be only a small number of events in the soft band;
thus, the source will have a low S/N and so not meet the criteria for the
piled-up fit to be performed. But, those few counts will nonetheless show
a hole in the center of the PSF and the non-piled-up fit will give an
inaccurate position. In order to properly handle such events, a list of
sources found to be piled up in the total band was supplied to the
processing for the other energy bands. Any source found in those bands
which lay within 20 pixels of a piled up source (50 pixels if $S<0.1$)
was assumed to be the piled up source, and thus the loss-function fit
was performed regardless of the S/N; the $L$ threshold required for such
sources to be recorded as piled up in the sub-bands was reduced to 2.5.
Despite this, there were still cases where pile up was not properly
identified in the sub-bands, and instead multiple faint, non-piled up
sources were reported. These were identified and handled during the
creation of the unique source list (Section~\ref{sec:merge}).

For all sources for which the loss function was fitted, regardless of whether it
was accepted as necessary, the best-fitting loss function parameters were
included in the catalog, along with \cstat\ with and without the loss function
and a note of whether the preferred fit was that with or without pile up.

As can be seen from the above description, in our software the loss
function was applied to the PSF, i.e.\ it affects only the events expected
from the source. In reality, the situation is more complex since there
will also be background events present, and pile up is related only to
the event rate, not the event origin; that is, the background should
also be suppressed by pile up, but the loss function definition does not
account for this. In fact, this issue is generally irrelevant because the
source is, by definition, extremely bright and the background is negligible
in comparison. The exception is for cases where $S\to0$, giving a hole
in the center of the PSF, as all events are migrated to unfeasible grades
or energies. In reality, there will be no events in the CCD center
because of pile up, however our PSF model will contain no \emph{source}
counts, but background events are still present. Since the hole is
symmetrical, and the fit will be dominated by those regions where source
counts are present, this problem can be discounted.

\subsection{Likelihood tests and flags}
\label{sec:Ltest}

In Paper I, we determined whether a detection corresponded to a real source by means of a
`likelihood' value (hereafter \lsrc) as defined in Equation~\ref{eq:L}, where
$\Delta\cstat$=$\cstat_{\rm nosrc}-\cstat_{\rm best}$: here $\cstat_{\rm best}$ is the fit
statistic of the PSF fit, and $\cstat_{\rm nosrc}$ is the C-stat value obtained comparing the
background map with the data over the source PSF fit region, i.e.\ the C-stat obtained if there
were no source present. For non-piled-up sources, $\Delta\nu$ is $3$, whereas for piled-up sources it is 7
(i.e.\ the number of free parameters in the PSF fit). Sources were assigned a quality flag based
on \lsrc\ (calibrated via simulations, see Section~\ref{sec:sims}). The limitation of this
statistic is that it can have a high value for reasons other than the presence of a point source:
for example, diffuse emission or imperfectly modeled stray light (Section~\ref{sec:sl}) may be
`better fitted' with a PSF-like distribution of counts than only with the underlying background,
despite there being no point source present in reality. We therefore introduced an extra test for
2SXPS, to supplement \lsrc. A model was fitted to reproduce a homogeneous elevation in count rate
in the fit region. This model had a single free parameter: the normalization. The C-stat for this
model ($\cstat_{\rm flat}$) was recorded and \lflat\ calculated via Equation~\ref{eq:L}, comparing
$\cstat_{\rm flat}$ with $\cstat_{\rm best}$. Low values of \lflat\ indicated that the PSF-like
count distribution offered little improvement over a homoegeneous disitrbution, i.e.\ the
`detection' was unlikely to be a point source.

As in 1SXPS, we defined three possible source flags: \emph{Good},
\emph{Reasonable} and \emph{Poor}\footnote{1SXPS also contained `bad'
sources with a very low likelihood of being real. We dropped this for
2SXPS.}, and like 1SXPS, these were defined such that the spurious
source contamination level was 0.3\%, 1\% and $\le$10\% in the \emph{Good},
\emph{Good+Reasonable} and full catalog samples respectively. However,
this time when determining the source flag both \lsrc\ and \lflat\ were
taken into account. The relationship between these likelihoods and
source flag depends on the exposure time. In Paper I we determined this
relationship based on the exposure in the image, which, because
vignetting in XRT is modest, was a viable approach. Due to the larger
stacked images in 2SXPS this is no longer viable, as exposure can vary
dramatically across the image due to the varying number of overlapping observations. The exposure time used in flag
determination in this work was thus the exposure time at the location of
the source. Additionally, the dependence on exposure time is really a
proxy for dependence on the background level. The $L$ thresholds in
Paper I were determined using simulated total-band images, and thus were
likely over-conservative for the soft, medium and hard bands, in which
the background level is naturally lower. For this
work, we instead determined the mean background levels from 1SXPS in each of the energy
bands as a fraction of that in the total energy band.  When calculating the exposure
to use in determination of a source's flag, the actual mean exposure time at
the source position was multiplied by this fraction. 

The relationship between $L$ values, exposure and assigned flag was calibrated via
the simulations described in Section~\ref{sec:sims}. As in Paper I, we found that the threshold
$L$ values depended on exposure time, as shown in Table~\ref{tab:Lthresh}. 
Sources flagged \emph{Good} by their likelihood values
were downgraded to \emph{Reasonable} if they lay within 30 pixels (71\arcsec)
of fitted stray light emission, or if the mean background in the source region
was above $10^{-3}$ ct s$^{-1}$; this latter case indicating that the detection
was likely to have arisen in an area heavily affected by the PSF wings of a bright source.
Such detections can be real sources, but the contamination rate in these cases will be higher than
in the simulation used to calibrate flag settings; we demoted such sources to keep
the \emph{Good} sample as pure as possible.

\begin{deluxetable}{cccc}
\tablecaption{The threshold likelihood values for the different detection
flags; both likelihoods must be above these for a source to be given the described flag.}
\tablehead{
\colhead{Flag}      &  \colhead{Exposure$^1$ range}   &     \colhead{\lsrc}  & \colhead{\lflat} 
} 
\startdata
\emph{Good}         &  $E\ge1000$ s      &  $18.293 E^{-0.0607}$  & 4 \\
                    &  $300\le E<1000$ s &   --- " ---            & 0 \\
                    &  $E<300$ s         &   14.8                 & 0 \\
\emph{Reasonable}   &  $E\ge1000$ s      &  $14.788 E^{-0.0562}$  & 6 \\
                    &  $300\le E<1000$ s &   --- " ---            & 0 \\
                    &  $E<300$ s         &   12.7                 & 0 \\
\emph{Poor}         &  $E\ge40000$ s     &  $7.7873 E^{-0.0433}$  & 6 \\
                    &  $300\le E<40000$ s&   --- " ---            & 0 \\
                    &  $E<300$ s         &   6.4                 & 0 \\
\enddata
\tablecomments{$^1E$ = exposure at the source position, scaled by the background
in the given band relative to the total band.}
\label{tab:Lthresh}
\end{deluxetable}

Within the database table, the flags are stored as integer values: 0, 1, and 2,  corresponding
to \emph{Good}, \emph{Reasonable} and \emph{Poor} respectively. These flags could be increased
to indicate concerns regarding the source. The extra values are bit-wise flags, described
in Table~\ref{tab:sflag}. So, for example, a source with a flag value of 5 would mean that the source
is \emph{Reasonable} (based on its likelihood values) but corresponds to a position covered by a known
extended source; thus it may be a point source within the extended emission, or it may be a spurious event
arising due to the extended emission.

As well as the detection flag, three other flags were created for each
source. `StrayLightWarning' was 0 or 1, indicating whether the source
had been flagged as being affected by stray light
(defined above). `NearBrightSourceWarning', 
indicates whether the mean background level at the source location was
high and so the source may be spurious due to a nearby bright object (see above). A value of
0 indicates that this warning is not set, and 1 indicates that it is. A value of 2 can also be given
for sources detected in stacked images. In these cases, if there is a variable source which was briefly bright
and has been observed many times, the PSF wings in the stacked image will have a low overall count-\emph{rate},
and the time-averaged PSF model may underestimate the PSF wings. So for any source detected in a stacked image,
in which the background rate is high (i.e.\ above $10^{-3}$ ct s$^{-1}$ pixel$^{-1}$ as above) in any individual observation of the source's location, the 
`NearBrightSourceWarning' is set to 2; the flag associated with that detection is also downgraded
from \emph{Good} to \emph{Reasonable} if it was the former. Another flag, `OpticalLoadingWarning', indicates whether the source
was potentially affected by optical loading; that is, whether its
position matched a known optical source bright enough to deposit
sufficient energy onto the XRT to masquerade as X-rays\footnote{See \url{https://www.swift.ac.uk/analysis/xrt/optical\_loading.php}.}. If no such
optical source was found, this flag was set to 0, otherwise its value
indicates how many magnitudes brighter the optical source is than the
magnitute at which optical loading is first expected to be a factor.
Optical loading is discussed in more detail in Paper I, section~3.4.

\begin{deluxetable}{ccc}
\tablecaption{Definition of the  bits in the detection flag that were set to indicate a potential problem.}
\tablehead{
\colhead{Bit}      &       \colhead{Value}  & \colhead{Meaning}   \\
} 
\startdata
2   &  4  & Source is within the extent of a known extended source. \\
3   &  8  & Source likely a badly-fitted piled-up source$^1$. \\
4   &  16  & Position matched area flagged by visual screening$^2$. \\
\enddata
\tablecomments{$^1$ See Section~\ref{sec:merge}.\\
$^2$ See Section~\ref{sec:screen}.}
\label{tab:sflag}
\end{deluxetable}

\subsection{Visual screening and flagging of datasets}
\label{sec:screen}

After source detection had been completed in all four energy bands, a dataset
was flagged as needing manual examination if any of the following criteria were satisfied:

\begin{enumerate}
\item{A possible source of stray light had been found (regardless of whether it was deemed necessary in the fitting).}
\item{The median distance between detections in any given band was $<80\arcsec$.}
\item{The dataset corresponded to an observation in 1SXPS which had a non-zero flag after manual screening in that catalog.}
\item{The dataset was a stacked image, for which one of the component observations satisfied criteria 2 or 3 above.} 
\end{enumerate}

The first criterion required us to verify that the stray light modelling
was at least adequate in all cases. Criterion 2 was specified because a
high density of observations either indicated a genuinely dense field
(such as the core of M31), or the presence of an artifact that gave rise
to multiple spurious detections such as diffuse emission or unmodelled
stray light. Criterion 3 is self-explanatory, and criterion 4 ensured
that any stacked image containing a potentially contaminated
observation was also checked. In total 13,825 datasets (out of 142,064)
were identified in this way. For each of these, two questions were
addressed: whether there was any diffuse emission present in the
image and whether stray light was handled adequately. In the former
case, if diffuse emission was identified in the image we created
circular or elliptical region(s) to cover the emission. All sources
lying inside this region were flagged (bit 4 of their flag set, see
Section~\ref{sec:Ltest} and Table~\ref{tab:sflag}). If this emission was astrophysical (i.e.\ not
arising due to instrumental effects such or bright Earth contamination)
then the region was also applied to all other observations sharing a
target ID with the screened image, and to any stacked image the
observation in question contributed. 

For fields where stray light had been deemed unnecessary by the fit, we confirmed
that it was indeed absent. For observations where stray light had been fitted,
the stray light model was compared with the image to confirm that stray light
was indeed present, the model gave a reasonably accurate reconstruction of the
stray light and that any detection within the stray light was appropriately
flagged. In the event that stray light was present but not fitted, or was fitted
but badly, we manually generated stray light images for trial positions of the
causal source until a reasonable reproduction of the observed stray light was
obtained. The observation was then reanalysed, with the manually-determined
position provided as the starting point for the stray light fit. The field was
then reinspected to confirm that the stray light was now handled.
Some fields contained stray light which had not been modelled, because the
bright source responsible for the stray light was not present in the catalogs
we searched (Section~\ref{sec:sl}). These fields thus contained mny spurious detections
and were identified by criterion 2 above, and were handled in the same way as fields where the stray light was
badly fitted.

In some cases even after refitting, the model fit was clearly
imperfect (for example, the curvature of the rings was not quite right,
or an extra ring was modelled which was not seen in the data); however,
provided that the model had been able to suppress spurious detections,
or at least ensure that such detections were flagged, it was accepted as
`good enough'. In a small number of cases (104) even after this iteration
an acceptable reproduction of the stray light could not be obtained. For
such fields bit two of the `field flag' for the affected dataset was set, as
described below.

For observations where stray light had been modelled,
but visual inspection showed that there was in fact no such
contamination, the observation was reanalysed with no stray light fitted.
If this reanalysis resulted in a median inter-source distance of
$<80\arcsec$ we reinspected the field to confirm whether our decision to
remove the stray light model had been erroneous (in which case it was reinstated). If an observation had
to be reanalysed as a result of the stray light screening, all 
stacked images to which that observation contributed were also reanalysed.

Once visual screening was complete, each dataset was assigned a flag, referred to as the `field flag' in the catalog tables.
This is a bitwise flag, and the different flags are defined in Table~\ref{tab:fieldflag}.

\begin{deluxetable}{ccc}
\tablecaption{Definition of the field flag, assigned to each dataset.}
\tablehead{
\colhead{Bit}      &       \colhead{Value}  & \colhead{Meaning}   \\
} 
\startdata
0   &  1  & Stray light was present, and fitted. \\
1   &  2  & Diffuse emission identified. \\
2   &  4  & Stray light badly/not fitted. \\
3   &  8  & Bright source fitting issues$^1$ \\
\enddata
\tablecomments{$^1$ i.e.\ the field contained a source that was heavily piled up in one band, but not fitted as such
in another band. See Section~\ref{sec:merge} for details.}
\label{tab:fieldflag}
\end{deluxetable}

\subsection{Construction of the unique source list}
\label{sec:merge}

The rationalization of detections into a unique list of sources was a
two-step process: first the detections from the different bands within
each dataset were combined into a single source list per observation.
For this step, the astrometric uncertainty on the XRT pointing could be
neglected since it was the same for each band. The second step was to
combine the outputs of Step 1 from each dataset into a unique list of
sources; for this of course the uncertainty in the relative astrometry
of the different observations had to be accounted for.

For Step 1, two detections in different bands were considered the same
source if their positions agreed to within 3 $\sigma$ or they were within 10 pixels (23.6\arcsec) of each
other. The latter clause arose because our simulations showed a large
tail on the position reconstruction error. This differs notably from
1SXPS where the match radius was simply a function of source brightness.
The reference position for the source was taken from the \emph{Good} or \emph{Reasonable}
detection with the smallest position error or, if all detections were \emph{Poor},
the detection with the highest S/N. In the event of a detection
in one band matching multiple detections from other bands, it was
assigned to that to which it was closest.

The exception to this was for piled-up sources. As discussed in
Section~\ref{sec:pup}, it is possible for a heavily piled-up source not
to be identified as such in the sub-band images, giving rise to multiple
spurious detections. To ameliorate this issue the detection-matching
process was carried out for piled-up sources first. For each piled-up
detection, $d_i$, in a given band, $b$, a counterpart was sought in each
other energy band with a localisation within 5 pixels (11.8\arcsec) and
an $S$ value within 0.5 of that of $d_i$. If
such a detection was found in band $b$ it was associated with $d_i$;
if not, then the pile-up profile of source $d_i$ had not been
fitted properly in band $b$; therefore all sources in that band
within 100 pixels (236\arcsec) of $d_i$ were assumed to be spurious
detections of $d_i$. Such detections had bit 3 of their flag set
(Table~\ref{tab:sflag}).

Once the unique list of sources per dataset had been determined (in step one), the
absolute astrometry of the dataset was calculated by aligning these
sources with the 2MASS catalog (see Paper I, section~3.7 for technical
details). If this process was successful, the corrected positions were
reported, but they were only used if the uncertainty on the 2MASS-derived
astrometric solution was smaller than that associated with the XRT star
tracker astrometry (3.5\arcsec\ at 90\%\ confidence).

For Step 2, the unique source lists from each dataset were compared, and
objects were considered the same source if their position -- including
astrometric uncertainty -- agreed at the
5-$\sigma$\footnote{Specifically, at the probability associated with a
Gaussian 5-$\sigma$ confidence. Since the radial position errors should
follow a Rayleigh distribution, this level was determined based on
Rayleigh, not Gaussian, statistics. Note that the statistical part
of the errors used here was overestimated; see footnote~\ref{fn:errerr}, p\pageref{fn:errerr}.} level. The final detection flag
assigned to each source and band was the best flag from all the
individual detections in that band; the overall detection flag and S/N
for each source was the best obtained from all detections and bands. The
stray light and optical loading warnings for each source were set to the
worst values from the individual detections.

In a small number of cases multiple detections of the same source were erroneously recorded
as different sources, as their positions in the different detections differed by more than 5-$\sigma$, suggesting
either a high proper motion, or that the position errors have a larger tail than would be expected from pure
Rayleigh statistics. The latter case will occur if, for example, the astrometric solution related to a field
has degenerate solutions, as can occur if (for example) the number of reference stars is low.
We therefore identified any sources which were within 20\arcsec\ of each other and not identified in the same dataset,
and marked them as potential aliases of each other. 1,735 sources were identified in this way. Not all of these
are aliases: some will be spurious events around a bright source, and some genuinely nearby but distinct sources.
However, these possible aliases are marked to allow users to investigate more closely if they desire.

%\subsection{Miscellaneous modifications, caveats and notes}
%\subsection{Caveats and warnings}

\section{Source products}
\label{sec:prods}

For each source we determined the count rate for each energy band and observation covering the
source location, regardless of whether it was detected in that dataset. We measured these rates
both averaged over the observation and for each individual snapshot. A circular region centered on
the source position was used, with the radius set to that used when the source was PSF fitted, or
12 pixels (28.2\arcsec) if the source was not detected in the dataset under consideration. The
total counts in that region, $C$, was measured from the image, and the expected number of
background counts $B$ was taken from the final background map for the observation/snapshot. If the
source had been detected in the observation in question, the PSF model for the source was first
subtracted from the background map. If $C-B>100$ or either $C>1000$ or $B>1000$, standard
frequentist statistics were used to determine the number of source counts and its error; otherwise
the Bayesian approach of \cite{Kraft91} was used. As in Paper I, we calculated the 1-$\sigma$
confidence interval on all count rates. However, in addition for this work, we calculated the
3-$\sigma$ interval for all observations and bands in which the source had not been detected, and
for all snapshots. If the 3-$\sigma$ lower limit was 0, the source was flagged as undetected in
this dataset, and the 3-$\sigma$ upper limit was recorded as well as the 1-$\sigma$ confidence
interval. Note that a source which was not found by the source detection process in a given
dataset can nonetheless be reported as detected in the same dataset by this `retrospective' count rate calculation
approach; this is because the source detection is a blind process, whereas retrospective count rate measurement
is predicated on the knowledge that there \emph{is}\/ a source at that location, which makes it more
sensitive (i.e.\ one does not need to allow for the large number of trial positions). When
accessing the source light curves via the 2SXPS website, users can choose whether to define a
datapoint as a `detection' based on the blind search or the retrospective analysis, and whether
to retrieve 3-$\sigma$ upper limits or 1-$\sigma$ confidence intervals for non detections, giving
greater control than was possible in 1SXPS. In addition to these time-resolved count rates, a
single mean count rate per energy band was determined by summing $C$ and $B$ from all the
individual observations. The peak rate in each band was also recorded, determined as the
individual per-observation or per-snapshot rate with the highest 1-$\sigma$ lower limit.

All count rates above were corrected for vignetting, pile up and bad
columns or pixels on the CCD. This was done by summing the fitted PSF
model (with pile up, if appropriate) multiplied by the exposure map over
the circular extraction regions, then also integrating the theoretical
PSF to a radius of 150 pixels\footnote{i.e. effectively infinity.} multiplied by
the peak on-axis exposure time.  The ratio of these gives the correction factor by 
which the count rate and error were multiplied. Note that, for large stacked images
the fractional exposure towards the edges can be very small compared to the
peak exposure time, giving very large corrections. When calculating the mean count rate,
the correction factor was calculated as $\sum_i(C_i F_i)/\sum_i C_i$, where $C_i$ is just $C$ measured
from dataset $i$, and $F_i$ is the correction factor in that observation.

\begin{figure}
\begin{center}
\includegraphics[width=8.1cm]{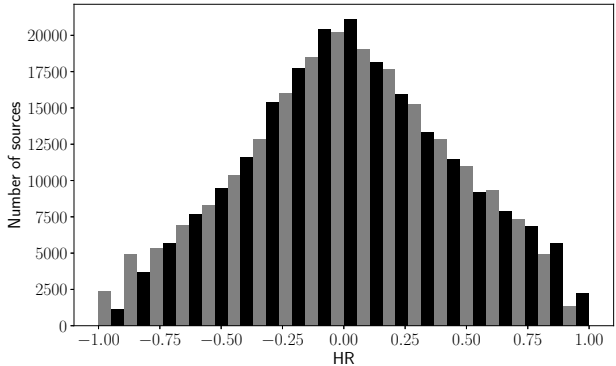}
\caption{The distribution of hardness ratio values from the sources in the catalog.
\emph{Gray}: HR1; \emph{Black}: HR2. The different colors are each half
the width of the actual bins.
}
\label{fig:HRdist}
\end{center}
\end{figure}

As well as light curves, two hardness ratios were created for each source,
for each snapshot and observation and an overall ratio. These ratios were defined as in Paper I:

\begin{equation}
\label{eq:HR1}
{\rm HR1} = (M-S)/(M+S)
\end{equation}

\begin{equation}
\label{eq:HR2}
{\rm HR2} = (H-M)/(H+M)
\end{equation}

\noindent where $S, M, H$ refer to the background-subtracted count-rates in the soft,
medium and hard bands respectively (the bands were defined in Table~\ref{tab:summary}). If both bands in the hardness ratio
contained $>100$ counts and had S/N$>2$ then the ratios were
calculated using the above equations, with the errors on $H$, $S$ and
$M$ taken as $\sqrt\{H,M,S\}$ respectively, and propagated through
Equations~\ref{eq:HR1} and \ref{eq:HR2}. For fainter sources we used the
Bayesian method of \cite{Park06}, where we used the effective area
option in their code to include the count rate correction factors in the
calculation. The distribution of HRs is shown in Fig.~\ref{fig:HRdist}.

For a small number of datasets with short exposures, there were no events in one or more
of the sub-bands, in which case the HRs could not be determined.

For each energy band and hardness ratio we also quantified source variability. 
This was done by creating per-snapshot and per-observation light curves
from the count rate and hardness ratios calculated as above; the 1-$\sigma$ confidence
intervals were used for all bins. The Pearson's \chisq\ \citep{Pearson00} was then
calculated as in Paper I, where the model was that of constant flux at the mean level,
and from this the probability that the source was constant was determined (see Paper I, section~4.1 
for details).

Note that, for all data products, we used only the PC mode data included
in the catalog. Many of the sources have also been covered by
WT mode observations. However, these contain only 1-D
spatial information and so are only appropriate for bright sources: with
the majority of 2SXPS sources being serendipitous, the WT data will be
contaminated by the other sources in the field\footnote{Normally WT mode
is only used for bright sources, where the number of photons from field
sources is negligible compared to those from the source.}.

As noted above, a very small fraction ($<0.8\%$) of the sources in the catalog are potential aliases
of other sources; in these cases the light curve will contain a mixture of correct source count rates
and erroneous measurements or upper limit (the latter in the case that the detections of the source
were classed under its alias). The 2SXPS website (Section~\ref{sec:contents}) allows the light curves of 
aliases to be combined in order to give the correct data.

\subsection{Spectral information and flux conversions}
\label{sec:spec}

Spectral and flux information was determined for every source. The approach is summarised briefly
here; for full details see section~4.2 of Paper I. These values were determined for two spectral
models: an absorbed power law, and an absorbed optically thin plasma model (APEC
\citealt{Smith01}); absorption was calculated using the {\sc tbabs} model \citep{Wilms00}. Flux
conversions and (where appropriate) spectral properties were derived using {\sc xspec}. Up to
three methods were used to determine the spectral details for each model.

The first method was applied to every source. We assumed standard emitting models: a power law
with photon index 1.7 and an APEC with a plasma temperature of 1 keV, and fixed the absorption
column to the Galactic value along the line of sight to the source, from \cite{Willingale13}. The
second method was attempted for every source. For this we simulated spectra in {\sc xspec} to
produce a look-up table of the spectral parameters (i.e.\ absorption column and either power-law
photon index or APEC temperature) as a function of (HR1,HR2). For each source whose time-averaged
(HR1,HR2) values were consistent with those producible by such a spectrum, we interpolated on this
grid to determine the spectral parameters (and uncertainties), and hence also the energy
conversion factor (ECF)\footnote{i.e.\ the conversion from detected counts, to source flux.}. The third
method was only carried out for sources from which more than 50 net counts were detected, and
which were detected (either in the blind searching or the restrospective count rate determination)
in at least one single-observation dataset. For the 23,326 sources meeting these criteria,
spectra were constructed using the tools from \cite{Evans09}, combining only those individual
datasets in which the source was detected (again, via either definition) -- this is to avoid
diluting the S/N in the spectrum by including periods of background-only data. In this case the
spectral models were fitted to the extracted spectra to give the best-fitting parameters. Fitting
was carried out on the unbinned data, minimising the $\mathcal{W}$ statistic in {\sc xspec}; a
Churazov-weighted \chisq\ \citep{Churazov96} was then calculated to give a goodness-of-fit
indication\footnote{Note that this \chisq\ cannot be used to calculate the null hypothesis
probability.}. We were able to obtain a fit with $\rchisq<1.2$ for
15,714 sources using the power-law model, and 12,314 using the APEC model. 11,444 sources yielded  $\rchisq<1.2$ for both models. 
Because the spectra
were built only from data where the source was detected, the fluxes given in the spectral
fits are biased. We therefore did not include these fluxes in the catalog; instead the ECF
for each source was derived from the spectral fit, and then multiplied by the mean total-band
count rate to give the flux. The count rate of each source was determined from all observations covering its
location, regardless of whether it was detected, and are thus not subject to this bias.

The only deviation of this approach from the Paper I method affected the second
method (HR interpolation). In Paper I we created a single set of look-up tables for each of the two
spectral models. However, on 2007 August 30 the CCD substrate voltage
was changed from 0V to 6V. This has a small effect on the spectral calibration
of the instrument, so for this work we created separate look-up tables for the two
substrate voltage settings. We chose which table to use based on whether the mean
arrival time for photons from the source occured before or after the voltage change.

For the APEC spectral model, there is a small region of (HR1, HR2) space which would be occupied by
sources with very high absorption columns ($>10^{22}$ \cms) and typically low ($<1$ keV) plasma temperatures;
for such sources the predicted counts to \emph{unabsorbed} flux conversion is very high (due to a low predicted
count rate, but high unabsorbed flux). There is a small number of sources, $<$1,000, in 2SXPS which thus
contain unrealistically high unabsorbed flux values, based on the interpolated APEC spectrum; such values should
be treated with caution, and are more likely to indicate that the true source spectrum is not an absorbed APEC.
The \emph{observed} fluxes for these objects are realistic, since these have, like the count rate, been
suppressed by the high absorption.

\section{Cross correlation with other catalogs}
\label{sec:xcorr}
\begin{deluxetable*}{ccrr}
%\begin{deluxetable}{cccc}
\tablecaption{Catalogs cross-correlated with 2SXPS.}
\tablehead{
\colhead{Catalog}      &       \colhead{Systematic Error$^1$}  & \colhead{Number of matches$^2$}  &\colhead{Spurious matches$^3$} \\
} 
\startdata
1SWXRT$^4$                      &            & 35,046  & 1,427    (4.1\%) \\
1SXPS$^5$                       &            & 98,378  & 3,223    (3.3\%) \\
2CSC$^6$                        &            & 9,273   & 602      (6.5\%) \\
2MASS$^7$                       &            & 73,707  & 43,222   (59\%) \\
2RXS$^8$                        &  25\arcsec & 11,447  & 1,433    (13\%) \\
3XMM DR8$^9$                    &            & 35,225  & 3,275    (9.3\%)\\
3XMM Stack$^{10}$               &            & 6,938   & 236      (3.4\%)\\
ALLWISE$^{11}$                  &            & 156,229 & 70,543   (45\%) \\
ROSHRI$^{12}$                   &  10\arcsec & 3,096   & 365      (12\%)\\
SDSS Quasar Catalog DR14$^{13}$ &            & 9,201   & 75       (0.9\%) \\
SwiftFT$^{14}$                  &            & 8,985   & 208      (2.3\%) \\
USNO-B1$^{15}$                  &            & 128,902 & 65,539   (51\%) \\
XMM SL2$^{16}$                  &  17\arcsec & 7,247   & 2,157    (30\%) \\
XRTGRB$^{17}$                   &            & 1,188   & 9        (0.8\%)\\
\enddata
\tablecomments{$^1$ 90\%\ confidence \newline $^2$ Number of 1SXPS sources for which there is a counterpart
in the external catalog within 3-$\sigma$. \newline $^3$ The number of 1SXPS sources with a match
after the 1SXPS position has been moved by 1--2\arcmin; the value in brackets is this number as a percentage of the matches to 1SXPS positions
for the same external catalog.
\newline 
$^4$    \cite{delia13};                                                                           % 1SWXRT
$^5$ \cite{Evans14}                                                                               % 1SXPS
$^6$    \cite{iEvans10};                                                                          % CSC
$^7$ \cite{Skrutskie06}                                                                           % 2MASS
$^8$ \cite{Boller16};                                                                             % 2RXS 
$^9$    \cite{Rosen16}, \url{http://xmmssc.irap.omp.eu/Catalogue/3XMM-DR8/3XMM\_DR8.html};        % 3XMM-DR8
$^{10}$ \cite{Traulsen19}                                                                         % 3XMM stack
$^{11}$ \url{http://wise2.ipac.caltech.edu/docs/release/allwise/}                                 % ALLWISE
$^{12}$ \url{http://heasarc.gsfc.nasa.gov/W3Browse/rosat/roshri.html}                             % ROSHRI
$^{13}$    \cite{Paris18};                                                                        % SDSS QSO
$^{14}$    \cite{puccetti11};                                                                     % SwiftFT    
$^{15}$ \cite{Monet03};                                                                           % USNO0B1    
$^{16}$  \cite{saxton08};                                                                         % XMM SL (?1/2)    
$^{17}$    Taken from \url{http://www.swift.ac.uk/xrt\_positions}; see \cite{Evans09};            % XRTGRB
}
\label{tab:xcorr}
\end{deluxetable*}

We cross-correlated the 2SXPS catalog with a range of other catalogs, using the same
approach as Paper I (section 4.3), i.e.\ identifying all sources in those catalogs with positions
agreeing with the 2SXPS position at the 99.7\%\ confidence level (using Rayleigh statistics, accounting for the uncertainty
in the 2SXPS and external catalogs\footnote{The statistical part of the 2SXPS error was slightly overestimated; see footnote~\ref{fn:errerr}, p\pageref{fn:errerr}}). Unlike Paper I
we chose not to correlate against the dynamic catalogs of SIMBAD and NED (links to perform such a search are
provided on the 2SXPS website), but we added correlations with ALLWISE and 1SXPS. For the other catalogs
we used updated versions if they existed; the list of catalogs and number of matches are given in Table~\ref{tab:xcorr}.
As for 1SXPS, we estimated the rate of spurious correlations by randomly shifting the 2SXPS positions by 1--2\arcmin\ and
repeating the correlation: the number of matches found in this second pass is also shown in  Table~\ref{tab:xcorr}.

\section{Catalog characteristics, access and contents}
\label{sec:contents}

2SXPS contains 206,335 unique sources, with a total of 1.1 million blind
detections across all four energy bands\footnote{In the \xmm\ catalogues
the detection of the same source in multiple energy bands in the same
dataset counts as a single detection. Using this terminology, 2SXPS
contains 530,612 detections.}. The median 0.3--10 keV
flux\footnote{Assuming an absorbed power-law spectrum.} is 4.7\tim{-14}
erg \cms\ s$^{-1}$. The observations in the catalog contain a total of
267 Ms, with a unique sky coverage of 3,790 deg$^2$. This is nearly
twice as much sky area as was covered by 1SXPS, 3.5 times the area
covered by 3XMM-DR8, and 6.8 times that in CSC
2.0\footnote{\url{http://cxc.harvard.edu/csc/char.html}}. There are
82,324 variable sources\footnote{Variable with at least 3-$\sigma$
significance.} in the catalog. Despite the lower effective area of XRT
compared to the \xmm\ instruments, the median source flux is only a
factor of two higher than in 3XMM-DR8, likely due to the lower
background level in XRT caused by its low-Earth orbit. 

The median source flux is higher than in 1SXPS, despite the fact that
our improved source detection system is actually more sensitive (Section~7).
This results from the combination of two effects. The first is a
result of our different data selection criteria compared to. 1SXPS. The other
factor is a result of the significant evolution of \swift\ science
operations over the past several years, as we have moved to more and
more short, wide surveys for galactic point sources and neutrino
and gravitational wave counterparts. The result of these
changes is a mean observation time of 2063 s in 2SXPS, compared to 3007
s in 1SXPS.

The catalog can be queried or downloaded via a dedicated website at: {\bf
https://www.swift.ac.uk/2SXPS}. Four tables are available for download, containing the sources and
their properties, individual detection details, details of the datasets in the catalog, and
details of the external catalog cross-correlation. The contents of these tables are described in
Appendix~\ref{sec:app_tabs}, Tables~\ref{tab:sourceDesc}--\ref{tab:xcorrDesc}. The main table, detailing the unique sources,
is also is available through Vizier, as catalog IX/58, and will be made available through HEASARC. 

The source and dataset tables can be queried via the above website, either using a simple cone search
or using detailed filtering on any/all of the table properties. Web pages exist for each source
and dataset, giving access to all products. An upper limit service is also provided. Full documentation is
on the website.

As for 1SXPS, we have defined a set of filters defining a `clean' sample, and additionally for 2SXPS an `ultra-clean' sample.
Cone searches on the website can be restricted to these subsamples. Clean sources are those with 
a best detection flag of 0 or 1 (i.e. \emph{Good} or \emph{Reasonable} with no other warning bits set);
OpticalLoadingWarning, StrayLightWarning and NearBrightSourceWarning all unset; and a field flag of 0 
or 1 (see Table~\ref{tab:fieldflag}). Ultra-clean sources are  a subset of the clean sources, with detection and field flags of 0.
There are 146,768 clean sources and 132,287 ultra-clean sources in 2SXPS.

\section{Completeness, contamination and accuracy}
\label{sec:sims}

\begin{figure}
\begin{center}
\includegraphics[height=8.2cm,angle=-90]{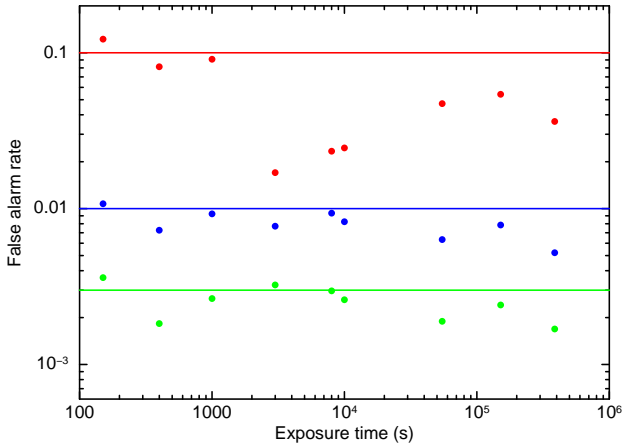}
\caption{The false positive rate from the simulations as a function of exposure time.
The solid lines are at the 0.3\%, 1\% and 10\% levels, and green, blue and red points
represent the \emph{Good}, \emph{Good + Reasonable} and complete catalog samples. For some exposure times
the false positive rate was never as high as the fiducial value for that flag, so those
contamination levels should be treated as conservative.}
\label{fig:false}
\end{center}
\end{figure}

% see https://www.facebook.com/groups/123898011017097/2967670486639821/?comment_id=2968567853216751&notif_id=1567771155573334&notif_t=group_comment 
% for notes on colours

\begin{figure}
\begin{center}
\includegraphics[height=8.2cm,angle=-90]{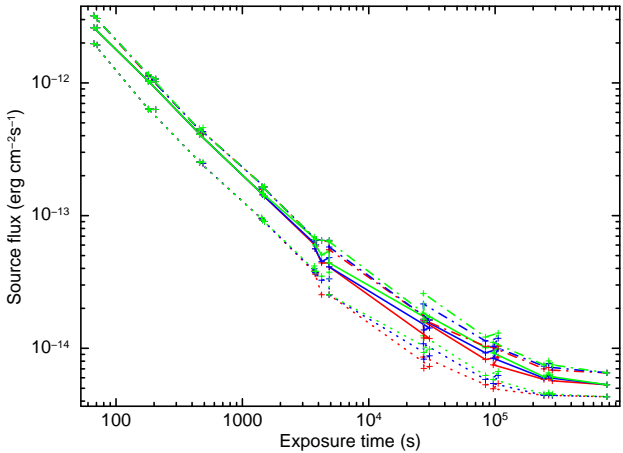}
\caption{The completeness of the 2SXPS catalog as determined from the simulations.
The dotted, solid and dot-dashed lines represent the flux at which 10\%, 50\%, and 90\%
of the simulated sources were recovered, as a function of exposure time.
The green, blue and red lines represent the \emph{Good}, \emph{Good + Reasonable}
and complete catalog samples respectively.}
\label{fig:complete}
\end{center}
\end{figure}

As for Paper I, we used simulations to calibrate the likelihood
thresholds and explore the performance of our source detection software.
We used the background maps (minus source models) from 1SXPS as the
input to the simulation; the background was modelled by randomly drawing
the number of counts in each pixel from a Poisson distribution with a
mean given by the background map. Sources were added to the image, with
their fluxes drawn randomly from the $\log N-\log S$ distribution of
extragalactic sources from \cite{Mateos08}. The number of sources per
image was also drawn at random from this distribution, although we
required a minimum of 10 sources per image to allow us to generate
reasonable statistics without requiring a ridiculous number of
simulations. We artificially spaced sources to be at least 50 pixels
apart, to ensure that the association of detected sources with simulated
sources was unambiguous; this may mean that the source completeness in
crowded fields is slightly less than from our simulations.

We simulated images with exposures approximately evenly distributed
(logarithmically) between 150 s and 1 Ms; for each exposure
time\footnote{Except for 1 Ms, where there was only one 1SXPS field
available.} we selected three seed datasets from 1SXPS, representing a
typical, low and high background level. We then simulated images; the
number of simulations performed depended on the exposure time since the
shorter images contained fewer sources, but were also quicker to
process: details are given in Table~\ref{tab:simInputs}. Our source
detection system was applied to these simulation. Detected sources were
either identified with one of the simulated sources (based on position and error), or marked
as spurious. We then calibrated the relationship between \lsrc, \lflat, exposure time and
detection flag, so as to maximise completeness while obtaining the false positive
rates for the different flags as defined in Section~\ref{sec:Ltest}; the resultant
thresholds were given in Table~\ref{tab:Lthresh}. Verification
of the false positive rate can be seen in Fig.~\ref{fig:false}. The completeness 
as a function of exposure time is shown in Fig.~\ref{fig:complete}. This represents a
significant increase in sensitivty over Paper I: in a 10 ks observation, the flux at which 2SXPS
is 50\%\ complete is 3.5 times lower than in 1SXPS. Note that, while we simulated based on three input datasets
for each exposure time, and the seed datasets did not have exactly identical exposures, for ease of viewing,
we have grouped each set into a single point in these figures. 

The reliability of the
count rate reconstruction (including effects of the Eddington bias; \citealt{Eddington40}), flux estimation
using the HRs, and variability estimates were all demonstrated in Paper I and we do not repeat that work here.

\begin{deluxetable}{cccc}
\tablecaption{The observations from 1SXPS used as inputs for our simulations.}
\tablehead{
\colhead{ObsID}      &  \colhead{Exposure}   &     \colhead{BG level$^1$}  & \colhead{Number of} \\
                     &                       &  & \colhead{simulations}             
} 
\startdata
00032223001	          &      150	s       &  6.15E-07   &   20,000 \\
00030051001	          &      150	s       &  8.46E-07   &   20,000 \\
00045199001	          &      150	s       &  1.31E-06   &   20,000 \\
00031189041	          &      399	s       &  7.80E-07   &   20,000 \\
00032433001	          &      399	s       &  1.45E-06   &   20,000 \\
00020001001	          &      401	s       &  5.46E-07   &   20,000 \\
00047148001	          &      1.0 ks	      &  5.84E-07   &   6,500 \\
00032200177	          &      1.0 ks	      &  7.36E-07   &   6,500 \\
00031468029	          &      1.0 ks	      &  1.73E-06   &   6,500 \\
00035306018	          &      3.0 ks       &  5.83E-07   &   3,500 \\
00031142001	          &      3.0 ks       &  7.47E-07   &   3,500 \\
00039846003	          &      3.0 ks       &  1.58E-06   &   3,500 \\
00037134002	          &      8.0 ks	      &  7.73E-07   &   1,000 \\
00040508003	          &      8.0 ks	      &  5.93E-07   &   1,000 \\
00051950063	          &      8.0 ks	      &  1.09E-06   &   1,000 \\
00037238001	          &      10.0 ks	    &  5.78E-07   &   1,000 \\
00232683000	          &      10.0 ks	    &  7.80E-07   &   1,000 \\
00416485007	          &      10.0 ks	    &  1.05E-06   &   1,000 \\
00302506000	          &      54 ks	      &  5.23E-07   &   1,000 \\
Stacked im 7508	      &      55 ks	      &  1.11E-06   &   1,000 \\
00163136014	          &      55 ks	      &  6.86E-07   &   1,000 \\
Stacked im 7133	      &      150 ks	      &  6.42E-07   &   1,000 \\
Stacked im 7130	      &      150 ks	      &  7.78E-07   &   1,000 \\
Stacked im 7616	      &      153 ks	      &  1.35E-06   &   1,000 \\
Stacked im 5470	      &      360 ks	      &  6.24E-07   &   1,000 \\
Stacked im 7005	      &      400 ks	      &  7.90E-07   &   1,000 \\
Stacked im 7032	      &      405 ks	      &  8.30E-07   &   1,000 \\
Stacked im 7086	      &      1.2 Ms   	  &  7.08E-07   &   1,000 \\
\enddata
\tablecomments{$^1$ i.e. the mean level in the source-less 1SXPS background map
in counts s$^{-1}$ pixel$^{-1}$.}
\label{tab:simInputs}
\end{deluxetable}

\section{Next steps}

\swift\ observes a large number of fields every day, and over recent years
this observation rate has increased: 2SXPS contains 2.6 times as many observations
as 1SXPS, yet only covers 1.5 times as much clock time (163 months, compared to 107 months).
The combination of large sky coverage and good source sensitivity makes the SXPS catalogs
a valuable reference to use when identifying possible X-ray transients. For example,
when searching for counterparts to gravitational wave triggers many uncatalogued X-ray sources
may be found and it is important to know whether they are new transient events or old sources in an 
area of sky previously uncataloged to XRT-levels of sensitivity.  

Due to the delay between an observation being carried out and the data
being incorporated in a catalog release, rather than waiting some years
and then producing 3SXPS, we are instead intending to produce a `live`
\emph{Swift}-XRT Point Source catalog (LSXPS) which will be updated each
time a new observation is completed. This will also be a powerful
facility for searching for transients or outbursts of known events in
real time.  This project is in its nascent stages at the moment; however,
we anticipate issuing periodic static catalog releases (3SXPS, 4SXPS)
to provide a resuable and fixed reference, but these will simply be
time-frozen snapshots of LSXPS.

\newpage

\section{Catalog usage}

This catalog can be freely used, provided this paper is cited; we also ask users to include
the following text in the acknowledgments of any paper using 2SXPS: \emph{This work made use of data supplied by the UK 
Swift Science Data Centre at the University of Leicester.}

\acknowledgments

PAE, KLP, JPO, and APB acknowledge UKSA support.
This research has made use of the XRT Data Analysis Software (XRTDAS) developed under the
responsibility of the ASI Space Science Data Center (ASDC), Italy. We made use of the GNU Scientific
Libraries (\url{https://www.gnu.org/software/gsl}; \citealt{GSL}); the {\sc Minuit2} package
provided by CERN
(\url{http://project-mathlibs.web.cern.ch/project-mathlibs/sw/Minuit2/html/index.html}), the {\sc
nlopt} fitting library (\url{https://github.com/stevengj/nlopt}) and {\sc xspec}. This research
has made use of data obtained from the Chandra Source Catalog, provided by the Chandra X-ray
Center (CXC) as part of the Chandra Data Archive. This research has made use of data obtained from
the 3XMM XMM-Newton serendipitous source catalogue compiled by the 10 institutes of the XMM-Newton
Survey Science Centre selected by ESA.
This publication makes use of data products from the Wide-field Infrared Survey Explorer, which is
a joint project of the University of California, Los Angeles, and the Jet Propulsion
Laboratory/California Institute of Technology, funded by the National Aeronautics and Space
Administration.

We are particularly indebted to Eric Mandel, developer of JS9 (\url{https://js9.si.edu/}), for his
support during the creation of the 2SXPS website.

\bibliographystyle{aasjournal}
\bibliography{phil}

\begin{thebibliography}{}
\expandafter\ifx\csname natexlab\endcsname\relax\def\natexlab#1{#1}\fi
\providecommand{\url}[1]{\href{#1}{#1}}
\providecommand{\dodoi}[1]{doi:~\href{http://doi.org/#1}{\nolinkurl{#1}}}
\providecommand{\doeprint}[1]{\href{http://ascl.net/#1}{\nolinkurl{http://ascl.net/#1}}}
\providecommand{\doarXiv}[1]{\href{https://arxiv.org/abs/#1}{\nolinkurl{https://arxiv.org/abs/#1}}}

\bibitem[{{Adri{\'a}n-Mart{\'{\i}}nez}
  {et~al.}(2016){Adri{\'a}n-Mart{\'{\i}}nez}, {Ageron}, {Albert}, {Samarai},
  {Andr{\'e}}, {Anton}, {Ardid}, {Aubert}, {Baret}, {Barrios-Mart{\'{\i}}},
  {Basa}, {Bertin}, {Biagi}, {Bogazzi}, {Bormuth}, {Bou-Cabo}, {Bouwhuis},
  {Bruijn}, {Brunner}, {Busto}, {Capone}, {Caramete}, {Carr}, {Chiarusi},
  {Circella}, {Coniglione}, {Costantini}, {Coyle}, {Creusot}, {Dekeyser},
  {Deschamps}, {De Bonis}, {Distefano}, {Donzaud}, {Dornic}, {Drouhin},
  {Dumas}, {Eberl}, {Els{\"a}sser}, {Enzenh{\"o}fer}, {Fehn}, {Felis},
  {Fermani}, {Folger}, {Fusco}, {Galat{\`a}}, {Gay}, {Gei{\ss}els{\"o}der},
  {Geyer}, {Giordano}, {Gleixner}, {Gracia-Ruiz}, {Graf}, {van Haren},
  {Heijboer}, {Hello}, {Hern{\'a}ndez-Rey}, {Herrero}, {H{\"o}{\ss}l},
  {Hofest{\"a}dt}, {Hugon}, {James}, {de Jong}, {Kadler}, {Kalekin}, {Katz},
  {Kie{\ss}ling}, {Kooijman}, {Kouchner}, {Kreykenbohm}, {Kulikovskiy},
  {Lahmann}, {Lambard}, {Lattuada}, {Lef{\`e}vre}, {Leonora}, {Loucatos},
  {Mangano}, {Marcelin}, {Margiotta}, {Mart{\'{\i}}nez-Mora}, {Martini},
  {Mathieu}, {Michael}, {Migliozzi}, {Moussa}, {Mueller}, {Neff}, {Nezri},
  {P{\u a}v{\u a}la{\v s}}, {Pellegrino}, {Perrina}, {Piattelli}, {Popa},
  {Pradier}, {Racca}, {Riccobene}, {Richter}, {Roensch}, {Rostovtsev},
  {Salda{\~n}a}, {Samtleben}, {Sanguineti}, {Sapienza}, {Schmid}, {Schnabel},
  {Schulte}, {Sch{\"u}ssler}, {Seitz}, {Sieger}, {Spurio}, {Steijger},
  {Stolarczyk}, {S{\'a}nchez-Losa}, {Taiuti}, {Tamburini}, {Trovato},
  {Tselengidou}, {T{\"o}nnis}, {Turpin}, {Vallage}, {Vall{\'e}e}, {Van
  Elewyck}, {Vecchi}, {Visser}, {Vivolo}, {Wagner}, {Wilms}, {Zornoza},
  {Z{\'u}{\~n}iga}, {Klotz}, {Boer}, {Le Van Suu}, {Akerlof}, {Zheng}, {Evans},
  {Gehrels}, {Kennea}, {Osborne}, \& {Coward}}]{Adrian-Martinez16}
{Adri{\'a}n-Mart{\'{\i}}nez}, S., {Ageron}, M., {Albert}, A., {et~al.} 2016,
  \jcap, 2, 062, \dodoi{10.1088/1475-7516/2016/02/062}

\bibitem[{{Arnaud}(1996)}]{Arnaud96}
{Arnaud}, K.~A. 1996, in Astronomical Society of the Pacific Conference Series,
  Vol. 101, Astronomical Data Analysis Software and Systems V, ed. G.~H.
  {Jacoby} \& J.~{Barnes}, 17

\bibitem[{{Boller} {et~al.}(2016){Boller}, {Freyberg}, {Tr{\"u}mper}, {Haberl},
  {Voges}, \& {Nandra}}]{Boller16}
{Boller}, T., {Freyberg}, M.~J., {Tr{\"u}mper}, J., {et~al.} 2016, \aap, 588,
  A103, \dodoi{10.1051/0004-6361/201525648}

\bibitem[{{Burrows} {et~al.}(2005){Burrows}, {Hill}, {Nousek}, {Kennea},
  {Wells}, {Osborne}, {Abbey}, {Beardmore}, {Mukerjee}, {Short}, {Chincarini},
  {Campana}, {Citterio}, {Moretti}, {Pagani}, {Tagliaferri}, {Giommi},
  {Capalbi}, {Tamburelli}, {Angelini}, {Cusumano}, {Br{\"a}uninger}, {Burkert},
  \& {Hartner}}]{BurrowsXRT}
{Burrows}, D.~N., {Hill}, J.~E., {Nousek}, J.~A., {et~al.} 2005, \ssr, 120,
  165, \dodoi{10.1007/s11214-005-5097-2}

\bibitem[{{Calabretta} \& {Greisen}(2002)}]{Calabretta02}
{Calabretta}, M.~R., \& {Greisen}, E.~W. 2002, \aap, 395, 1077,
  \dodoi{10.1051/0004-6361:20021327}

\bibitem[{{Cash}(1979)}]{Cash79}
{Cash}, W. 1979, \apj, 228, 939, \dodoi{10.1086/156922}

\bibitem[{{Churazov} {et~al.}(1996){Churazov}, {Gilfanov}, {Forman}, \&
  {Jones}}]{Churazov96}
{Churazov}, E., {Gilfanov}, M., {Forman}, W., \& {Jones}, C. 1996, \apj, 471,
  673, \dodoi{10.1086/177997}

\bibitem[{{D'Elia} {et~al.}(2013){D'Elia}, {Perri}, {Puccetti}, {Capalbi},
  {Giommi}, {Burrows}, {Campana}, {Tagliaferri}, {Cusumano}, {Evans},
  {Gehrels}, {Kennea}, {Moretti}, {Nousek}, {Osborne}, {Romano}, \&
  {Stratta}}]{delia13}
{D'Elia}, V., {Perri}, M., {Puccetti}, S., {et~al.} 2013, \aap, 551, A142,
  \dodoi{10.1051/0004-6361/201220863}

\bibitem[{{Ebisawa} {et~al.}(2003){Ebisawa}, {Bourban}, {Bodaghee}, {Mowlavi},
  \& {Courvoisier}}]{Ebisawa03}
{Ebisawa}, K., {Bourban}, G., {Bodaghee}, A., {Mowlavi}, N., \& {Courvoisier},
  T.~J.-L. 2003, \aap, 411, L59, \dodoi{10.1051/0004-6361:20031336}

\bibitem[{{Eddington}(1940)}]{Eddington40}
{Eddington}, Sir, A.~S. 1940, \mnras, 100, 354

\bibitem[{{Evans} {et~al.}(2010){Evans}, {Primini}, {Glotfelty}, {Anderson},
  {Bonaventura}, {Chen}, {Davis}, {Doe}, {Evans}, {Fabbiano}, {Galle}, {Gibbs},
  {Grier}, {Hain}, {Hall}, {Harbo}, {(Helen He}, {Houck}, {Karovska},
  {Kashyap}, {Lauer}, {McCollough}, {McDowell}, {Miller}, {Mitschang},
  {Morgan}, {Mossman}, {Nichols}, {Nowak}, {Plummer}, {Refsdal}, {Rots},
  {Siemiginowska}, {Sundheim}, {Tibbetts}, {Van Stone}, {Winkelman}, \&
  {Zografou}}]{iEvans10}
{Evans}, I.~N., {Primini}, F.~A., {Glotfelty}, K.~J., {et~al.} 2010, \apjs,
  189, 37, \dodoi{10.1088/0067-0049/189/1/37}

\bibitem[{{Evans} {et~al.}(2009){Evans}, {Beardmore}, {Page}, {Osborne},
  {O'Brien}, {Willingale}, {Starling}, {Burrows}, {Godet}, {Vetere}, {Racusin},
  {Goad}, {Wiersema}, {Angelini}, {Capalbi}, {Chincarini}, {Gehrels}, {Kennea},
  {Margutti}, {Morris}, {Mountford}, {Pagani}, {Perri}, {Romano}, \&
  {Tanvir}}]{Evans09}
{Evans}, P.~A., {Beardmore}, A.~P., {Page}, K.~L., {et~al.} 2009, \mnras, 397,
  1177, \dodoi{10.1111/j.1365-2966.2009.14913.x}

\bibitem[{{Evans} {et~al.}(2014){Evans}, {Osborne}, {Beardmore}, {Page},
  {Willingale}, {Mountford}, {Pagani}, {Burrows}, {Kennea}, {Perri},
  {Tagliaferri}, \& {Gehrels}}]{Evans14}
{Evans}, P.~A., {Osborne}, J.~P., {Beardmore}, A.~P., {et~al.} 2014, \apjs,
  210, 8, \dodoi{10.1088/0067-0049/210/1/8}

\bibitem[{{Evans} {et~al.}(2015){Evans}, {Osborne}, {Kennea}, {Smith},
  {Palmer}, {Gehrels}, {Gelbord}, {Homeier}, {Voge}, {Strotjohann}, {Cowen},
  {Boeser}, {Kowalski}, \& {Stasik}}]{Evans15}
{Evans}, P.~A., {Osborne}, J.~P., {Kennea}, J.~A., {et~al.} 2015, \mnras, 448,
  2210, \dodoi{10.1093/mnras/stv136}

\bibitem[{{Evans} {et~al.}(2016{\natexlab{a}}){Evans}, {Osborne}, {Kennea},
  {Campana}, {O'Brien}, {Tanvir}, {Racusin}, {Burrows}, {Cenko}, \&
  {Gehrels}}]{Evans16}
---. 2016{\natexlab{a}}, \mnras, 455, 1522, \dodoi{10.1093/mnras/stv2213}

\bibitem[{{Evans} {et~al.}(2016{\natexlab{b}}){Evans}, {Kennea}, {Palmer},
  {Bilicki}, {Osborne}, {O'Brien}, {Tanvir}, {Lien}, {Barthelmy}, {Burrows},
  {Campana}, {Cenko}, {D'Elia}, {Gehrels}, {Marshall}, {Page}, {Perri},
  {Sbarufatti}, {Siegel}, {Tagliaferri}, \& {Troja}}]{Evans16c}
{Evans}, P.~A., {Kennea}, J.~A., {Palmer}, D.~M., {et~al.} 2016{\natexlab{b}},
  \mnras, 462, 1591, \dodoi{10.1093/mnras/stw1746}

\bibitem[{{Evans} {et~al.}(2017){Evans}, {Cenko}, {Kennea}, {Emery}, {Kuin},
  {Korobkin}, {Wollaeger}, {Fryer}, {Madsen}, {Harrison}, {Xu}, {Nakar},
  {Hotokezaka}, {Lien}, {Campana}, {Oates}, {Troja}, {Breeveld}, {Marshall},
  {Barthelmy}, {Beardmore}, {Burrows}, {Cusumano}, {D'A{\`i}}, {D'Avanzo},
  {D'Elia}, {de Pasquale}, {Even}, {Fontes}, {Forster}, {Garcia}, {Giommi},
  {Grefenstette}, {Gronwall}, {Hartmann}, {Heida}, {Hungerford}, {Kasliwal},
  {Krimm}, {Levan}, {Malesani}, {Melandri}, {Miyasaka}, {Nousek}, {O'Brien},
  {Osborne}, {Pagani}, {Page}, {Palmer}, {Perri}, {Pike}, {Racusin}, {Rosswog},
  {Siegel}, {Sakamoto}, {Sbarufatti}, {Tagliaferri}, {Tanvir}, \&
  {Tohuvavohu}}]{Evans17}
{Evans}, P.~A., {Cenko}, S.~B., {Kennea}, J.~A., {et~al.} 2017, Science, 358,
  1565, \dodoi{10.1126/science.aap9580}

\bibitem[{{Galassi} {et~al.}(2009){Galassi}, {Davies}, {Theiler}, {Gough},
  {Jungman}, {Alken}, {Booth}, \& {Rossi}}]{GSL}
{Galassi}, M., {Davies}, J., {Theiler}, J., {et~al.} 2009, GNU Scientific
  Library Reference Manual - Third Edition, 3rd edn. (Network Theory Ltd.)

\bibitem[{{Gehrels} {et~al.}(2004){Gehrels}, {Chincarini}, {Giommi}, {Mason},
  {Nousek}, {Wells}, {White}, {Barthelmy}, {Burrows}, {Cominsky}, {Hurley},
  {Marshall}, {M{\'e}sz{\'a}ros}, {Roming}, {Angelini}, {Barbier}, {Belloni},
  {Campana}, {Caraveo}, {Chester}, {Citterio}, {Cline}, {Cropper}, {Cummings},
  {Dean}, {Feigelson}, {Fenimore}, {Frail}, {Fruchter}, {Garmire}, {Gendreau},
  {Ghisellini}, {Greiner}, {Hill}, {Hunsberger}, {Krimm}, {Kulkarni}, {Kumar},
  {Lebrun}, {Lloyd-Ronning}, {Markwardt}, {Mattson}, {Mushotzky}, {Norris},
  {Osborne}, {Paczynski}, {Palmer}, {Park}, {Parsons}, {Paul}, {Rees},
  {Reynolds}, {Rhoads}, {Sasseen}, {Schaefer}, {Short}, {Smale}, {Smith},
  {Stella}, {Tagliaferri}, {Takahashi}, {Tashiro}, {Townsley}, {Tueller},
  {Turner}, {Vietri}, {Voges}, {Ward}, {Willingale}, {Zerbi}, \&
  {Zhang}}]{GehrelsSwift}
{Gehrels}, N., {Chincarini}, G., {Giommi}, P., {et~al.} 2004, \apj, 611, 1005,
  \dodoi{10.1086/422091}

\bibitem[{{Greisen} \& {Calabretta}(2002)}]{Greisen02}
{Greisen}, E.~W., \& {Calabretta}, M.~R. 2002, \aap, 395, 1061,
  \dodoi{10.1051/0004-6361:20021326}

\bibitem[{{IceCube Collaboration}(2013)}]{IceCube13}
{IceCube Collaboration}. 2013, Science, 342, 1242856,
  \dodoi{10.1126/science.1242856}

\bibitem[{{Kennea} {et~al.}(2018){Kennea}, {Coe}, {Evans}, {Waters}, \&
  {Jasko}}]{Kennea18}
{Kennea}, J.~A., {Coe}, M.~J., {Evans}, P.~A., {Waters}, J., \& {Jasko}, R.~E.
  2018, \apj, 868, 47, \dodoi{10.3847/1538-4357/aae839}

\bibitem[{{Kraft} {et~al.}(1991){Kraft}, {Burrows}, \& {Nousek}}]{Kraft91}
{Kraft}, R.~P., {Burrows}, D.~N., \& {Nousek}, J.~A. 1991, \apj, 374, 344,
  \dodoi{10.1086/170124}

\bibitem[{{Mateos} {et~al.}(2008){Mateos}, {Warwick}, {Carrera}, {Stewart},
  {Ebrero}, {Della Ceca}, {Caccianiga}, {Gilli}, {Page}, {Treister}, {Tedds},
  {Watson}, {Lamer}, {Saxton}, {Brunner}, \& {Page}}]{Mateos08}
{Mateos}, S., {Warwick}, R.~S., {Carrera}, F.~J., {et~al.} 2008, \aap, 492, 51,
  \dodoi{10.1051/0004-6361:200810004}

\bibitem[{{Monet} {et~al.}(2003){Monet}, {Levine}, {Canzian}, {Ables}, {Bird},
  {Dahn}, {Guetter}, {Harris}, {Henden}, {Leggett}, {Levison}, {Luginbuhl},
  {Martini}, {Monet}, {Munn}, {Pier}, {Rhodes}, {Riepe}, {Sell}, {Stone},
  {Vrba}, {Walker}, {Westerhout}, {Brucato}, {Reid}, {Schoening}, {Hartley},
  {Read}, \& {Tritton}}]{Monet03}
{Monet}, D.~G., {Levine}, S.~E., {Canzian}, B., {et~al.} 2003, \aj, 125, 984,
  \dodoi{10.1086/345888}

\bibitem[{{Moretti} {et~al.}(2004){Moretti}, {Campana}, {Tagliaferri}, {Abbey},
  {Ambrosi}, {Angelini}, {Beardmore}, {Br{\"a}uninger}, {Burkert}, {Burrows},
  {Capalbi}, {Chincarini}, {Citterio}, {Cusumano}, {Freyberg}, {Giommi},
  {Hartner}, {Hill}, {Mori}, {Morris}, {Mukerjee}, {Nousek}, {Osborne},
  {Short}, {Tamburelli}, {Watson}, \& {Wells}}]{Moretti04}
{Moretti}, A., {Campana}, S., {Tagliaferri}, G., {et~al.} 2004, Society of
  Photo-Optical Instrumentation Engineers (SPIE) Conference Series, Vol. 5165,
  {SWIFT XRT point spread function measured at the Panter end-to-end tests},
  ed. K.~A. {Flanagan} \& O.~H.~W. {Siegmund}, 232--240,
  \dodoi{10.1117/12.504857}

\bibitem[{{Moretti} {et~al.}(2005){Moretti}, {Campana}, {Mineo}, {Romano},
  {Abbey}, {Angelini}, {Beardmore}, {Burkert}, {Burrows}, {Capalbi},
  {Chincarini}, {Citterio}, {Cusumano}, {Freyberg}, {Giommi}, {Goad}, {Godet},
  {Hartner}, {Hill}, {Kennea}, {La Parola}, {Mangano}, {Morris}, {Nousek},
  {Osborne}, {Page}, {Pagani}, {Perri}, {Tagliaferri}, {Tamburelli}, \&
  {Wells}}]{Moretti05}
{Moretti}, A., {Campana}, S., {Mineo}, T., {et~al.} 2005, in Society of
  Photo-Optical Instrumentation Engineers (SPIE) Conference Series, Vol. 5898,
  UV, X-Ray, and Gamma-Ray Space Instrumentation for Astronomy XIV, ed.
  O.~H.~W. {Siegmund}, 360--368, \dodoi{10.1117/12.617164}

\bibitem[{{Moretti} {et~al.}(2009){Moretti}, {Pagani}, {Cusumano}, {Campana},
  {Perri}, {Abbey}, {Ajello}, {Beardmore}, {Burrows}, {Chincarini}, {Godet},
  {Guidorzi}, {Hill}, {Kennea}, {Nousek}, {Osborne}, \&
  {Tagliaferri}}]{Moretti09}
{Moretti}, A., {Pagani}, C., {Cusumano}, G., {et~al.} 2009, \aap, 493, 501,
  \dodoi{10.1051/0004-6361:200811197}

\bibitem[{{P{\^a}ris} {et~al.}(2018){P{\^a}ris}, {Petitjean}, {Aubourg},
  {Myers}, {Streblyanska}, {Lyke}, {Anderson}, {Armengaud}, {Bautista},
  {Blanton}, {Blomqvist}, {Brinkmann}, {Brownstein}, {Brand t}, {Burtin},
  {Dawson}, {de la Torre}, {Georgakakis}, {Gil-Mar{\'\i}n}, {Green}, {Hall},
  {Kneib}, {LaMassa}, {Le Goff}, {MacLeod}, {Mariappan}, {McGreer}, {Merloni},
  {Noterdaeme}, {Palanque-Delabrouille}, {Percival}, {Ross}, {Rossi},
  {Schneider}, {Seo}, {Tojeiro}, {Weaver}, {Weijmans}, {Y{\`e}che}, {Zarrouk},
  \& {Zhao}}]{Paris18}
{P{\^a}ris}, I., {Petitjean}, P., {Aubourg}, {\'E}., {et~al.} 2018, \aap, 613,
  A51, \dodoi{10.1051/0004-6361/201732445}

\bibitem[{{Park} {et~al.}(2006){Park}, {Kashyap}, {Siemiginowska}, {van Dyk},
  {Zezas}, {Heinke}, \& {Wargelin}}]{Park06}
{Park}, T., {Kashyap}, V.~L., {Siemiginowska}, A., {et~al.} 2006, \apj, 652,
  610, \dodoi{10.1086/507406}

\bibitem[{{Pearson}(1900)}]{Pearson00}
{Pearson}, K. 1900, Philosophical Magazine Series 5, 50, 157,
  \dodoi{10.1080/14786440009463897}

\bibitem[{{Popp} {et~al.}(2000){Popp}, {Hartmann}, {Soltau}, {Str{\"u}der},
  {Meidinger}, {Holl}, {Krause}, \& {von Zanthier}}]{Popp00}
{Popp}, M., {Hartmann}, R., {Soltau}, H., {et~al.} 2000, Nuclear Instruments
  and Methods in Physics Research A, 439, 567,
  \dodoi{10.1016/S0168-9002(99)00912-2}

\bibitem[{{Puccetti} {et~al.}(2011){Puccetti}, {Capalbi}, {Giommi}, {Perri},
  {Stratta}, {Angelini}, {Burrows}, {Campana}, {Chincarini}, {Cusumano},
  {Gehrels}, {Moretti}, {Nousek}, {Osborne}, \& {Tagliaferri}}]{puccetti11}
{Puccetti}, S., {Capalbi}, M., {Giommi}, P., {et~al.} 2011, \aap, 528, A122,
  \dodoi{10.1051/0004-6361/201015560}

\bibitem[{{Rosen} {et~al.}(2016){Rosen}, {Webb}, {Watson}, {Ballet}, {Barret},
  {Braito}, {Carrera}, {Ceballos}, {Coriat}, \& {Della Ceca}}]{Rosen16}
{Rosen}, S.~R., {Webb}, N.~A., {Watson}, M.~G., {et~al.} 2016, \aap, 590, A1,
  \dodoi{10.1051/0004-6361/201526416}

\bibitem[{Rowan(1990)}]{Rowan90}
Rowan, T.~H. 1990, PhD thesis, University of Texas at Austin, Austin, TX, USA

\bibitem[{{Saxton} {et~al.}(2008){Saxton}, {Read}, {Esquej}, {Freyberg},
  {Altieri}, \& {Bermejo}}]{saxton08}
{Saxton}, R.~D., {Read}, A.~M., {Esquej}, P., {et~al.} 2008, \aap, 480, 611,
  \dodoi{10.1051/0004-6361:20079193}

\bibitem[{{Shaw} {et~al.}(2017){Shaw}, {Heinke}, {Bahramian}, {Maccarone},
  {Kennea}, {Kuulkers}, {Degenaar}, {Sivakoff}, {Wijnands}, {Strader}, \& {in't
  Zand}}]{Shaw17}
{Shaw}, A.~W., {Heinke}, C.~O., {Bahramian}, A., {et~al.} 2017, in AAS/High
  Energy Astrophysics Division, Vol.~16, AAS/High Energy Astrophysics Division
  \#16, 400.01

\bibitem[{Singer {et~al.}(2014)Singer, Price, Farr, Urban, Pankow, Vitale,
  Veitch, Farr, Hanna, Cannon, Downes, Graff, Haster, Mandel, Sidery, \&
  Vecchio}]{Singer14}
Singer, L.~P., Price, L.~R., Farr, B., {et~al.} 2014, Astrophys. J., 795, 105,
  \dodoi{10.1088/0004-637X/795/2/105}

\bibitem[{{Skrutskie} {et~al.}(2006){Skrutskie}, {Cutri}, {Stiening},
  {Weinberg}, {Schneider}, {Carpenter}, {Beichman}, {Capps}, {Chester},
  {Elias}, {Huchra}, {Liebert}, {Lonsdale}, {Monet}, {Price}, {Seitzer},
  {Jarrett}, {Kirkpatrick}, {Gizis}, {Howard}, {Evans}, {Fowler}, {Fullmer},
  {Hurt}, {Light}, {Kopan}, {Marsh}, {McCallon}, {Tam}, {Van Dyk}, \&
  {Wheelock}}]{Skrutskie06}
{Skrutskie}, M.~F., {Cutri}, R.~M., {Stiening}, R., {et~al.} 2006, \aj, 131,
  1163, \dodoi{10.1086/498708}

\bibitem[{{Smith} {et~al.}(2001){Smith}, {Brickhouse}, {Liedahl}, \&
  {Raymond}}]{Smith01}
{Smith}, R.~K., {Brickhouse}, N.~S., {Liedahl}, D.~A., \& {Raymond}, J.~C.
  2001, \apjl, 556, L91, \dodoi{10.1086/322992}

\bibitem[{{Traulsen} {et~al.}(2019){Traulsen}, {Schwope}, {Lamer}, {Ballet},
  {Carrera}, {Coriat}, {Freyberg}, {Michel}, {Motch}, \& {Rosen}}]{Traulsen19}
{Traulsen}, I., {Schwope}, A.~D., {Lamer}, G., {et~al.} 2019, \aap, 624, A77,
  \dodoi{10.1051/0004-6361/201833938}

\bibitem[{{Vanderbilt} \& {Louie}(1984)}]{Vanderbilt84}
{Vanderbilt}, D., \& {Louie}, S.~G. 1984, Journal of Computational Physics, 56,
  259, \dodoi{10.1016/0021-9991(84)90095-0}

\bibitem[{{Voges} {et~al.}(1999){Voges}, {Aschenbach}, {Boller},
  {Br{\"a}uninger}, {Briel}, {Burkert}, {Dennerl}, {Englhauser}, {Gruber},
  {Haberl}, {Hartner}, {Hasinger}, {K{\"u}rster}, {Pfeffermann}, {Pietsch},
  {Predehl}, {Rosso}, {Schmitt}, {Tr{\"u}mper}, \& {Zimmermann}}]{Voges99}
{Voges}, W., {Aschenbach}, B., {Boller}, T., {et~al.} 1999, \aap, 349, 389

\bibitem[{{Watson} {et~al.}(2009){Watson}, {Schr{\"o}der}, {Fyfe}, {Page},
  {Lamer}, {Mateos}, {Pye}, {Sakano}, {Rosen}, {Ballet}, {Barcons}, {Barret},
  {Boller}, {Brunner}, {Brusa}, {Caccianiga}, {Carrera}, {Ceballos}, {Della
  Ceca}, {Denby}, {Denkinson}, {Dupuy}, {Farrell}, {Fraschetti}, {Freyberg},
  {Guillout}, {Hambaryan}, {Maccacaro}, {Mathiesen}, {McMahon}, {Michel},
  {Motch}, {Osborne}, {Page}, {Pakull}, {Pietsch}, {Saxton}, {Schwope},
  {Severgnini}, {Simpson}, {Sironi}, {Stewart}, {Stewart}, {Stobbart}, {Tedds},
  {Warwick}, {Webb}, {West}, {Worrall}, \& {Yuan}}]{Watson09}
{Watson}, M.~G., {Schr{\"o}der}, A.~C., {Fyfe}, D., {et~al.} 2009, \aap, 493,
  339, \dodoi{10.1051/0004-6361:200810534}

\bibitem[{Willingale(2019)}]{Willingale19}
Willingale, R. 2019, in Optics for EUV, X-Ray, and Gamma-Ray Astronomy IX, ed.
  S.~L. O'Dell \& G.~Pareschi, Vol. 11119, International Society for Optics and
  Photonics (SPIE), 149 -- 165, \dodoi{10.1117/12.2530682}

\bibitem[{{Willingale} {et~al.}(2013){Willingale}, {Starling}, {Beardmore},
  {Tanvir}, \& {O'Brien}}]{Willingale13}
{Willingale}, R., {Starling}, R.~L.~C., {Beardmore}, A.~P., {Tanvir}, N.~R., \&
  {O'Brien}, P.~T. 2013, \mnras, 431, 394, \dodoi{10.1093/mnras/stt175}

\bibitem[{{Wilms} {et~al.}(2000){Wilms}, {Allen}, \& {McCray}}]{Wilms00}
{Wilms}, J., {Allen}, A., \& {McCray}, R. 2000, \apj, 542, 914,
  \dodoi{10.1086/317016}

\bibitem[{{Wolter}(1952)}]{Wolter52}
{Wolter}, H. 1952, Annalen der Physik, 445, 94,
  \dodoi{10.1002/andp.19524450108}

\end{thebibliography}

\appendix
\section{Stray light modelling}
\label{sec:app_sl}

\begin{figure*}
\begin{center}
\includegraphics[width=14cm]{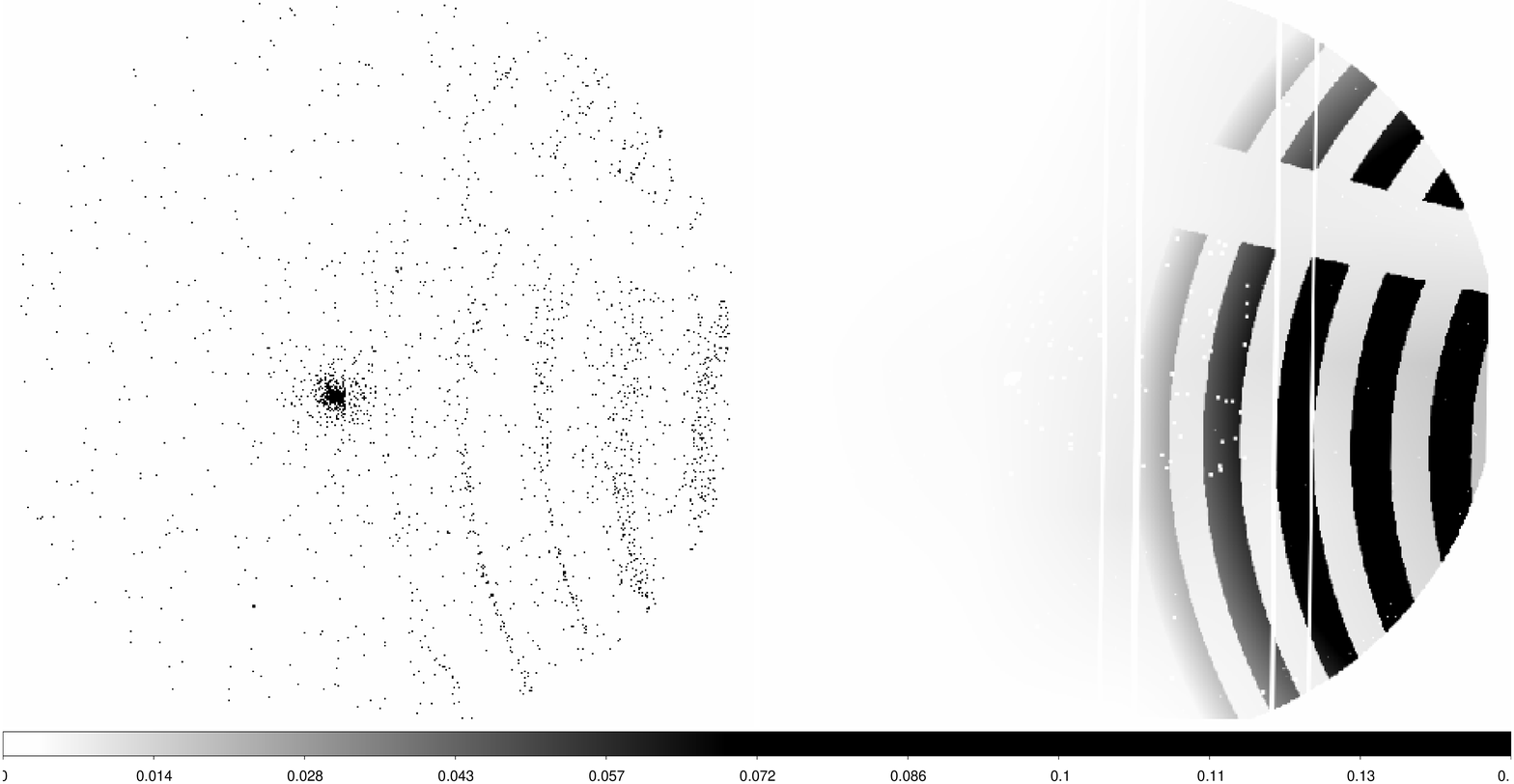}
\includegraphics[width=14cm]{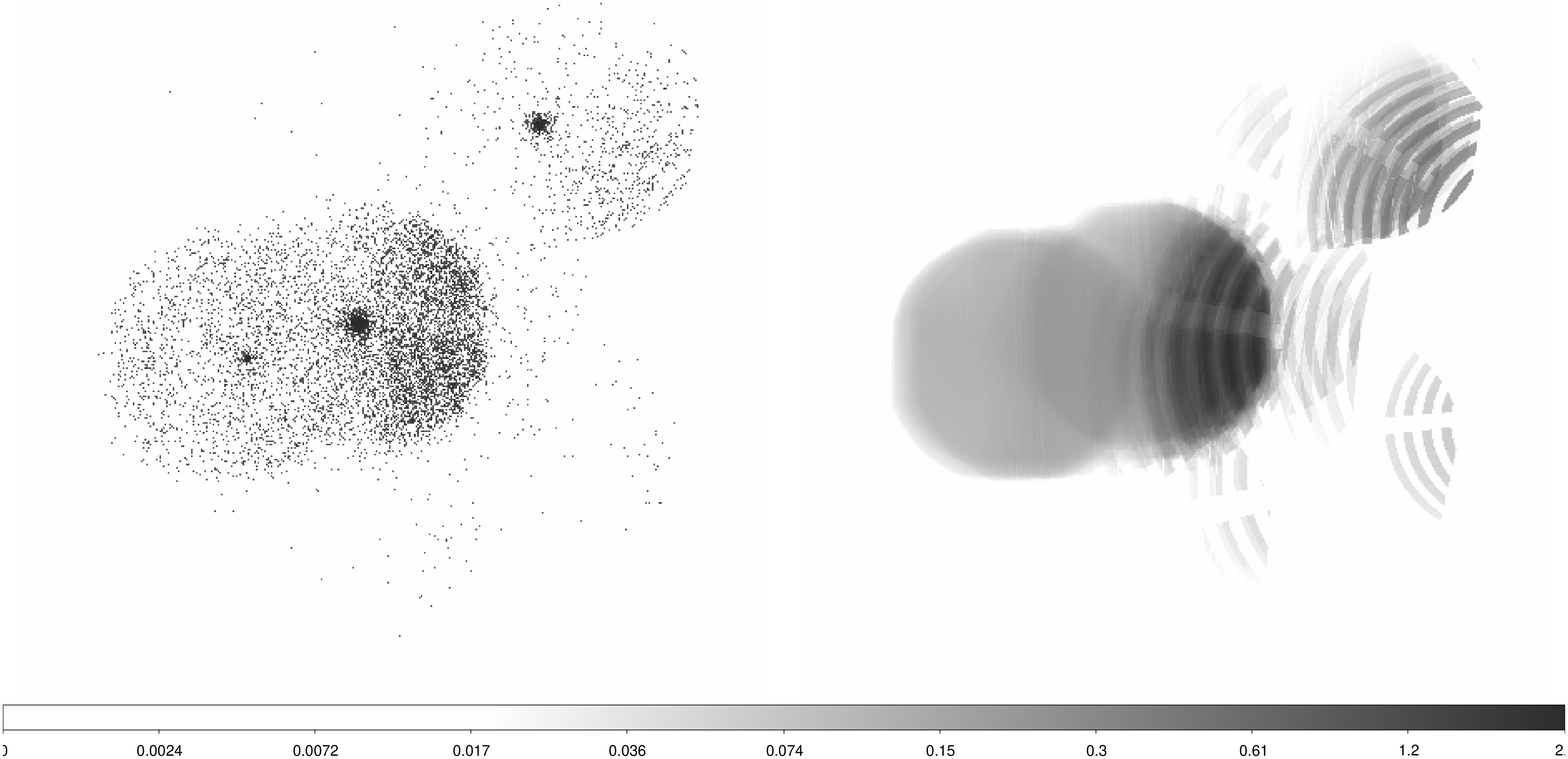}
\caption{Examples of stray light and its model. \emph{Top}: the data (left) and background map
(right) from obsid 00591551000. This dataset contains only a single shapshot of data, so the individual rings
are clear. \emph{Bottom}: As the top but for stacked image 14459, which includes the observation
from the top panel. A range of obsids are present in this stacked image, many of which suffer
stray light contamination from the same source (1SXPS J181228.2-181236). Where obsids have multiple snapshots
the effect of the different pointings can be seen as the stray light models overlap, and the shadows
caused by the mirror support structure move. The vertical stripes and white spots
are the result of the dead zones on the CCD from hot pixels or columns.
}
\label{fig:sl}
\end{center}
\end{figure*}

Stray light is a result of the Wolter-I optical design of X-ray telescopes such as the \swift-XRT.
X-rays originating outside of the nominal field of view undergo a single reflection off the second 
(hyperbolic) mirror surface, which scatters them onto the detector. The result is a concentric ring
pattern on the detector as shown in Fig.~\ref{fig:sl}. Each ring represents reflections 
off a single mirror shell; the arc-shapes resulting because the X-rays have reflected off a range of
azimuthal angles around the mirror, and the ring thickness arising from the extent along the mirror
length which can scatter the X-rays onto the CCD.

\cite{Willingale19} describe in detail how the shape of this
pattern can be determined for a given off-axis source position and the geometry
of the reflecting surface. Their model was originally produced for the Wide Field Imager
instrument on the forthcoming \emph{Athena} satellite, but is applicable to all
nested Wolter-I telescopes, such as \swift-XRT, and \xmm. To produce a model for stray light in XRT, 
we used the equations 
from Willingale \etal, with details of the XRT mirror
from the JET-X design specification\footnote{The mirrors on XRT were originally
fabricated for JET-X.}, which inluded the dimensions, shape and thickness of the
mirror shells, baffles and mirror support structure. We then used this model to
predict the stray light pattern on the XRT detector in terms of 
three input parameters: the position angle of the causal source relative to the
CCD $x$ axis ($\theta$), the off-axis angle of the source (i.e. the angle
between the CCD boresight and the source, $\phi$) and a flux normalization
($N$). The brightness of the rings was calculated using the X-ray reflectivity
of the mirrors, which depends on both photon energy and grazing angle.
Note that this model returned the number of counts expected in each CCD
pixel as a decimal, i.e. it is not quantised; it therefore served as a model to
which the real (quantised and Poisson-distributed) stray light detected could be
compared. To perform this comparison, the model image was converted from the CCD
detector frame to a sky-coordinate image, using the satellite pointing
information in a manner analogous to that used to convert the original event
lists into sky images.
% Although this is itself not an enormously arduous task by
% modern standards (applying the translation to 360,000 pixels), it had to be done
% for each snapshot independently (since the spacecraft pointing and roll angle
% changes), and when fitting stray light the runtime taken up by the translation
% was considerable. Fortunately, stray light never extends over the entire CCD, so
% we were able to determine a sub-box within the CCD where stray light was present
% and only consider pixels within this for translation. Similarly, pixels to which
% the stray light model contributed no events could be ignored. This provided a
% substantial speed up.

An example of the stray light model, converted to sky coordinates, is shown in Fig.~\ref{fig:sl},
along with the actual 0.3--10 keV image. As can readily be seen, the broad features of the data
are well reproduced by the model, however there are imperfections: the radius of curvature of the
rings is not quite right, and the radial intensity profiles are flatter and wider than the real
data. These arise because our model assumes the idealised mirror exactly as per the design, whereas
the real mirror has imperfections. The incorrect curvature arises because our model assumes that the XRT
mirror shells are perfectly circular in cross section, whereas in reality they are distorted slightly
by their connection to the mirror support structure. The radial profile differs from reality because
in the idealised model, each mirror shell is perfectly uniform in thickness, and the shells are exactly concentric
(i.e.\ the inter-shell spacing is constant); in the real mirror there are deviations from this
idealised scenario which alters radial profile of the rings.
A side-effect of the latter problem is that, while the total number of counts in the stray light
models was correct, the peak level in the center of the rings was underestimated and so the detection of 
spurious sources was not adequately reduced. We therefore increased the normalization 
of the stray light rings by 1.5 compared to that expected from the mirror model (this number reached by trial
and improvement). This has the side
effect of causing the background to be even more grossly overestimated at the edges of the rings,
although in fact this helps to compensate for the curvature errors. Pragmatically, our goal
was to suppress the detection of spurious sources resulting from stray light and to flag
any detected sources which were likely to be either spurious or at least affected by 
stray light; the fact that this approach may tend to over-estimate
the stray light is prefereable to the alternative.

\subsection{Incorporating stray light into the background model}
\label{sec:app_slfit}

\begin{figure*}
\begin{center}
\includegraphics[width=18cm]{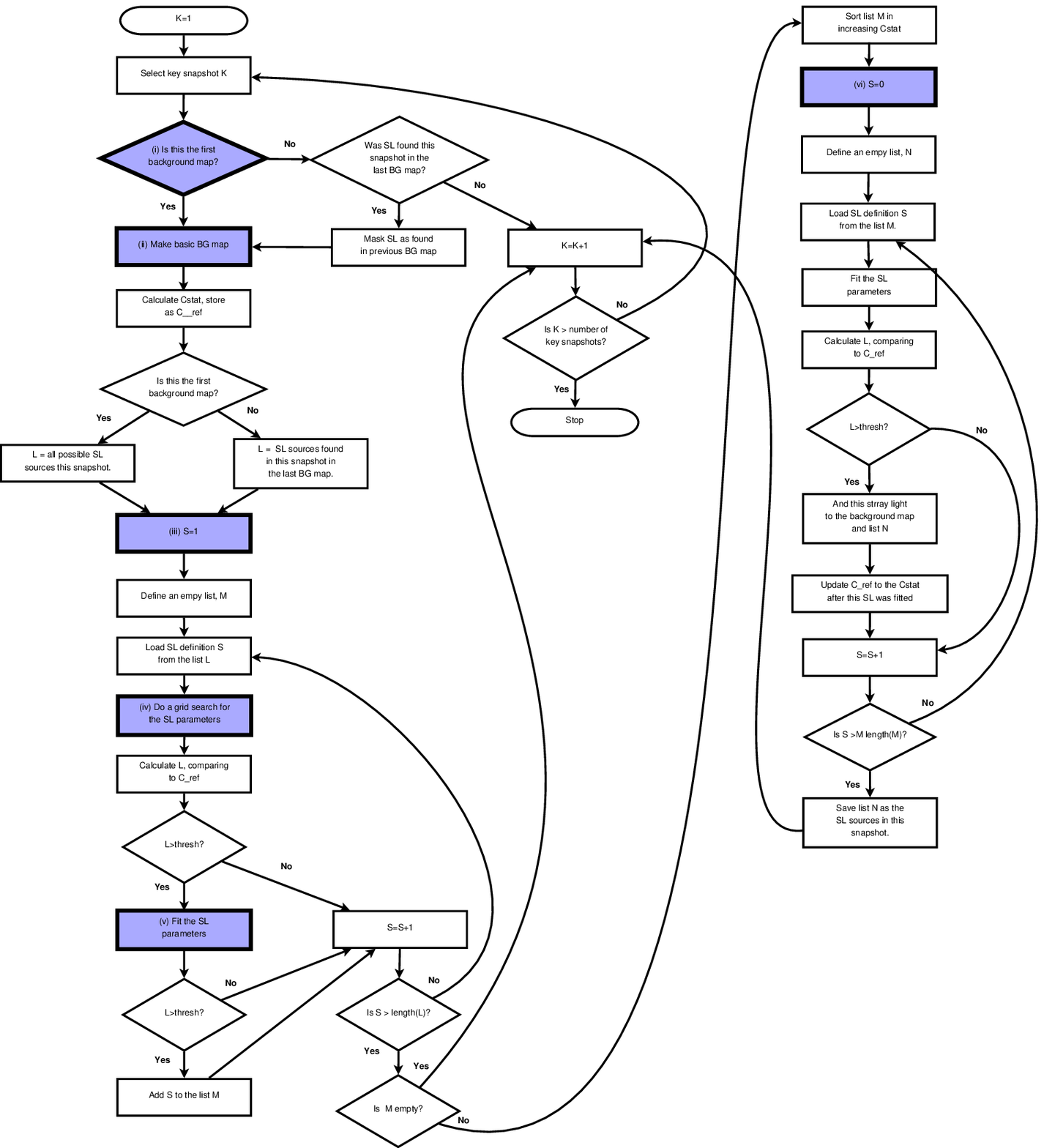}
\caption{Flow-chart depicting the algorithm used for searching for stray light, and including
a model of it in the background map. Lilac boxes with heavy borders mark the reference points numbered (lower-case Roman
numerals) in the text.}
\label{fig:slAlgo}
\end{center}
\end{figure*}

When analyzing a dataset containing stray light, the stray light model had to be fitted
to the dataset for the reasons discussed in Section~\ref{sec:slmodel}. This was a complex process,
illustrated in Fig.~\ref{fig:slAlgo} and described below.

Before any source detection or background modelling was carried out, the
snapshots were organized into co-pointed groups. Any snapshot pointed
within 80 pixels (3.1\arcmin) of an earlier snapshot was assigned to
the same group as that earlier one\footnote{If a snapshot lay within 80
pixels of multiple disjointed snapshots, it was assigned to whichever
group it was closest to.}. Within each group the snapshot with the
longest exposure (and so expected to have the best-sampled stray light)
was identified; these will be referred to hereafter as `key snapshots'.
During the actual background map creation (below), the full fit and test
of whether stray light was needed was carried out only for the key
snapshots; for the other snapshots, only the normalization was fitted:
the position was fixed. This was primarily for reasons of computational
efficiency: the fitting process was CPU intensive and slow; thus, by
reducing the number of snapshots for which the full fit was needed, the
overall runtime could be significantly reduced.

Not all steps in the fitting process (Fig.~\ref{fig:slAlgo}) were
carried out each time the background map was constructed, as indicated
by the decision forking. Here we describe the essential algorithm, with
the deviations from it explained afterwards. Note that this presupposes
that (a) potential source(s) of stray light had been identified as
described above (Section~\ref{sec:sl}); if not, none of the
stray-light-specific steps described here were carried out. Some steps
are indicated with lower-case Roman numerals below and in
Fig.~\ref{fig:slAlgo}, for ease of reference later on. 

The construction of the background map, as in Paper I, consisted of
iterating over all snapshots, creating a map per snapshot, and then
summing them. For datasets with possible stray light, the key snapshots
were processed first. 

%%NOTE: Theta is alpha in the code :/ but can't be alpha because that's confusing with RA.
%% Phi is phi in the code.
%% Theta is the angle relative to the x-axis
%% Phi is the off-axis angle

{\bf(i)} For each key snapshot, as well as masking out any
sources already detected, the regions of the CCD covered by stray light,
as modelled last time the background map was created, were also masked
out. {\bf(ii)} The `basic map' (i.e. that created by the mask/rebin/interpolate
approach) was then created. 
%%
%% CHECK THIS - it appears to be commented out in the code, so maybe I ended up screening everything!
% In this case not all detected sources were masked out,
% only those which had been PSF fitted and were found to have $L_{\rm flat}\ge<25$ (see
% Section~\ref{sec:Ltest}). This is because, until the stray-light has been adequately modelled, spurious
% sources are detected as a result of the stray light: masking all of these out would result
% in the stray light itself being masked out and so unfittable. Spurious sources caused by 
% stray light tend to have lower $L_{\rm flat}$ values than real sources, hence only point-like 
% objects were screened at this stage. 
%%
%% CONFIRMED FROM GIT COMMIT MESSAGES AND TIMES. I MADE THIS CHANGE TO MASK OUT EVEN LFLATS ON JUNE 18 2018, IE BEFORE RUNNING
%%
The C-stat was calculated (Equation~\ref{eq:cstat})
comparing this background map with the snapshot image data, this value was recorded
as $\cstat_{\rm ref}$.

{\bf(iii)} The possible stray light sources were then considered
independently. The first time the background map was created, the
positions of the stray-light sources were converted from ($\alpha,
\delta$) to ($\theta, \phi$), these being the parameters to be fitted
and stored internally. $\theta$, the position angle from the CCD $x$
axis to the source, was allowed to vary by $\pm5\deg$; $\phi$, the
off-axis angular distance to the source was given a range
$\pm10\arcmin$. The normalization was fitted in log space, and allowed
to vary by $\pm3$ dex from the initial estimated value (determined from the
cataloged flux of the source). {\bf(iv)} Initially a grid search was
performed to determine the best starting point for a fit. The three
parameters were stepped over their ranges in 5 steps, \cstat\ calculated
at each point, at the best parameters and \cstat\ noted. A likelihood
test (Equation~\ref{eq:L}) was carried out comparing this best \cstat\
with $\cstat_{\rm ref}$ determined in step (ii) and unless $L$ was at
least 15, stray light was deemed not to be present from this source and
it was ignored. {\bf(v)} For cases where $L\ge15$, a fit was performed,
using the best parameters from the grid search as the starting point,
but retaining the parameter limits from step (iii). Fitting was carried
out using the {\sc nlopt}
library\footnote{\url{https://github.com/stevengj/nlopt}} and the
{\sc nl\_sbplx} algorithm, which is based on the `subplex' algorithm of
\cite{Rowan90}. $L$ was calculated comparing \cstat\ from the best fit
with $\cstat_{\rm ref}$, and if $L\ge30$ the stray light source was
saved as a \emph{possible} contributor to stray light in this snapshot.

{\bf(vi)} Once steps (iii)--(v) had been performed for each possible
stray light source, any which passed the likelihood test were sorted
into decreasing order of fit quality (i.e. increasing order of \cstat).
These were again fitted as in step (v), except that this time the
likelihood threshold was increased to 32. If a stray light source passed
this threshold, it was deemed to be present in the data. The best-fitting
model of the stray light was immediately  added to the
background map used to fit the next possible stray light source, and $\cstat_{\rm
ref}$ was set to the \cstat\ value found from the fit. Thus, once a
stray light source had been identified a subsequent source could only
also be added to a key snapshot if it was still deemed significant given
the presence of the source(s) already identified. This was necessary
because even a badly-fitting or unnecessary stray light model gave a significant improvement
to \cstat\ when the true source of the stray light is not included in the
model.

{\bf(vii)} For any sources of stray light which passed all of the above steps,
the best fitting position parameters ($\theta, \phi$) were compared to
the best-fitting values from the last time a background map was created.
If the new position represented a shift in ($\alpha, \delta$) of
30\arcsec\ or more (or if this was the first background map, so no
previously fitted stray light positions existed), it is likely that the
stray light was incorrectly masked out during step (iii). So the entire
process [steps (iii)--(vi)] was repeated, using these new positions as the
starting point, and for masking. Note that all stray light sources that
had been tested during steps (iii)--(vi) were included again in this
pass, including those where $L$ had been found to be below threshold; this
is in case the improved masking changed the $L$ values.

After the above process had been carried out for each key snapshot, and
the resultant stray light sources and their parameters stored, the
remaining snapshots were processed. For these, the stray light
definitions were taken from the relevant key snapshot, masked out in the
creation of the basic map, and then fitted using the same library as
above [step (v)], but this time only the normalization was free to vary
and only by $\pm2$ dex. No likelihood tests were performed: all stray
light sources accepted for the key snapshot were modelled in each
snapshot within that group.

The above algorithm describes the overall approach followed each time a
background map was created during the source detection process, however
there were deviations from this approach. The list of possible
stray-light sources used from step (iii) onwards was not constant. In
the first background map, all possible stray light sources identified when the data
were being prepared were considered; in subsequent maps, only sources which passed the likelihood tests in
steps (iv)--(vi) were used in step (iii) of the next background map.
During the very first background map creation, the thresholds used in
steps (iv)--(vi) were all set to one initially; i.e. any possible stray
light source that made even a marginal improvement to the background
model was initially retained -- this was because at this point no stray
light had been masked in creation of the basic map, which could
significantly reduce the $L$ values returned. However, once step (vii)
was reached even on the first background map, the $L$ thresholds were
restored to those described above.

During source detection, once all of the high S/N sources had been detected and the S/N threshold
reduced (i.e. phase two, middle column of Fig.~\ref{fig:algorithm}, had begun) the
positions and number of stray light sources was fixed completely;
hereafter the key snapshots were handled like the normal snapshots,
i.e.\ only the normalization was able to be refitted.

The full stray-light fitting process described above was only carried
out on the total (0.3--10 keV) band image, and only for individual
observations. In the former case, this is the image with the most events
and so in which the stray light could be best modelled. Since the other
images are subsets of the total-band image, it is nonsensical to
independently fit the stray light; instead the positions of the stray
light sources (per snapshot) from the total image were supplied to the
source detection code for the soft, medium and hard bands and all
snapshots were treated as non-key snapshots, i.e.\ only the stray-light
normalization was fitted, and no likelihood tests were carried out.

For stacked images it was not necessary to carry out the full stray light
fitting, since the background mapping works on individual snapshots, 
regardless of what type of dataset is being analysed\footnote{In principle a stacked image could
yield more sources with S/N$>$ 10, which could have a small effect on the stray
light position. Such effect is small however, and the stray light
fitting is so computationally demanding that it is not sensible to run
it independently on the stacked image, for a negligible improvement.} Therefore,
for stacked images only the stray light normalization was fitted, and only
the stray light sources necessary for the component observations were used, with no
likelihood tests performed. 

A shortcoming of our approach is that it will not pick up stray light
too faint to make a significant impact on an individual snapshot, but
which is visible -- and produces spurious detections -- in the full
image for the dataset. In reality, this situation is rare, and was
caught during the visual screening phase (Section~\ref{sec:screen}). The only alternative
would be to simultaneously fit all snapshots, which is not practical.

\section{PSF calibration}
\label{sec:app_psf}

\begin{figure}
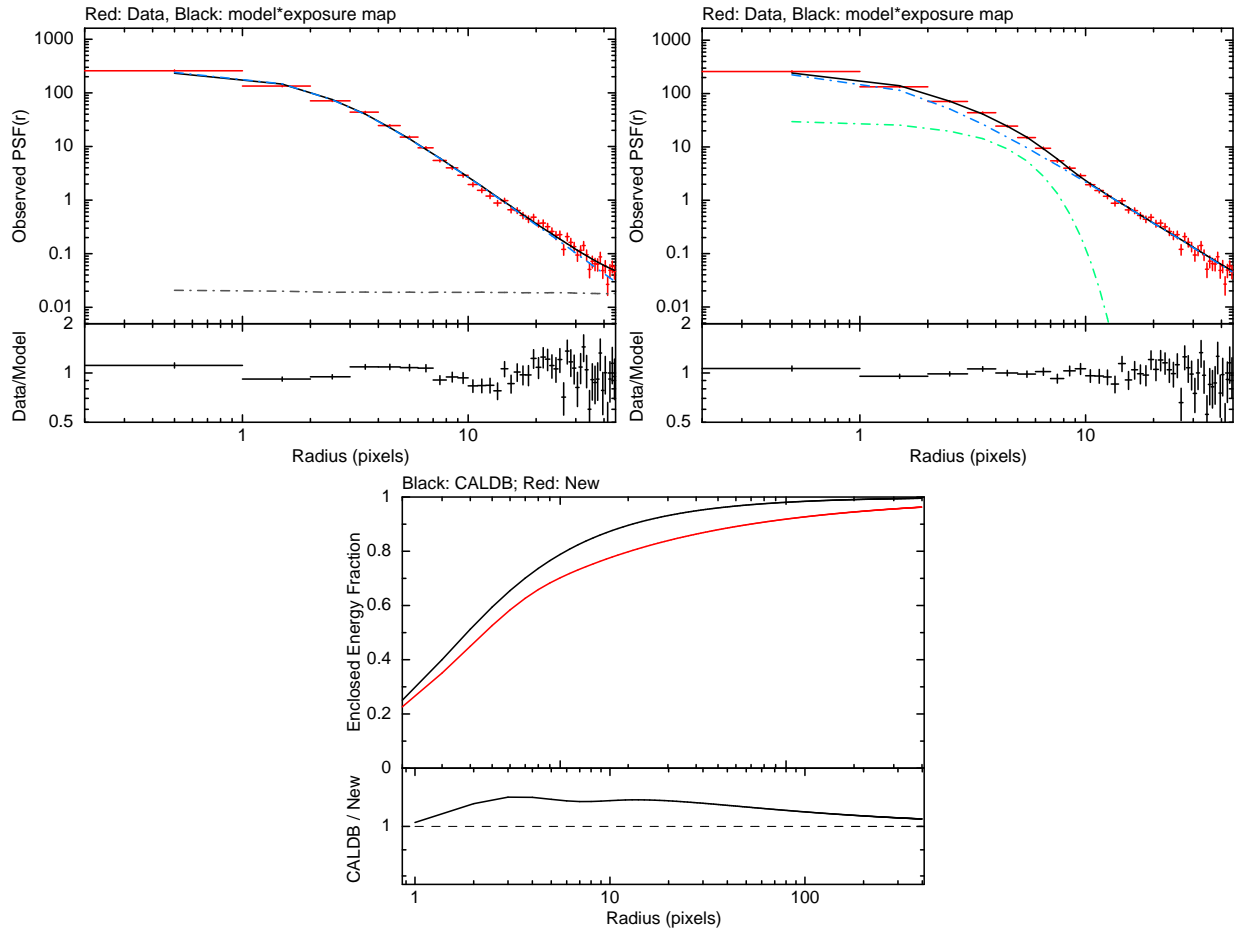

\begin{center}
\includegraphics[height=8.1cm,angle=-90]{Field1834_king.eps}
\includegraphics[height=8.1cm,angle=-90]{Field1834_kingGaus.eps}
\includegraphics[height=8.1cm,angle=-90]{compare_PSF.eps}
\caption{The CALDB (top-left) PSF model and our new PSF model (top-right) fitted to the same dataset. The blue, green and 
black dashed lines show the Gaussian and King components and fitted background. The improvement
in residuals around 8--15 pixels can clearly be seen. With the CALDB PSF, the counts beyond \til30 pixels
are all background counts; the new PSF has broader wings, which reproduce the events out to large radii
(the background is below the y-axis in this plot). \emph{bottom}: The enclosed energy fraction of the CALDB (black) and new (red)
PSF models, with the ratio of the two in the bottom panel.}
\label{fig:psfmodel}
\end{center}
\end{figure}

Calibrating the PSF wings (i.e.\ the regions more than \til30 pixels from the source) is
challenging, since these contain only a small fraction of the source flux. Bright
sources cannot be used for this calibration as their PSF shape is distorted by pile up. Instead
one must identify modest-brightness sources with long exposures. This is also problematic for
\swift\ because it has a low-Earth orbit; therefore, long exposures can only be achieved by
comibining data from multiple spacecraft pointings. The star trackers on-board \swift\ give
astrometric accuracy of 3.5\arcsec\ (90\% confidence), compared to a pixel size of 2.357\arcsec,
thus when combining the data one must account for the fact that the source position can move
slightly between observations, which will both broaden the PSF and change its shape. So, we
require sources bright enough for a sub-pixel localisation to be performed for each snapshot.

We therefore selected sources in the 1SXPS catalog with total-band count rates
in the range 0.3--0.6 s$^{-1}$, a minimum of five separate observations in the catalog,
and a Galactic absorption column below 3\tim{21} \cms. The latter criterion was to reduce the
risk of high foreground dust contamination which can broaden the PSF by scattering.
For each source we identified the stacked image in 1SPXS it was in, and used the data from that
to model the PSF. The data were split into snapshots, and we performed a source centroid on each snapshot 
individually, rejecting any snapshots where the 1-$\sigma$ position error was above 0.5 pixels.

We simultaneously fitted the same model to all snapshots
individually, where the source position for each snapshot was taken from the centroiding performed
above, as any form of shifting and adding the individual artificially broaden the PSF. 
We found that the fits tended to be prone to local minima and therefore fitted the PSF
profile using the simulated annealing approach of \cite{Vanderbilt84}. The fitted model was that
given in Equation~\ref{eq:psf}. The fits to some of the sources gave parameters significantly
discrepant from those obtained from the majority of sources, likely indicating either a failure to
find the true minimum, or possibly some issue with the data (such as a dust-scattering halo, or
contamination from an unresolved nearby source). We tried refitting with a slower `cooling rate'
for the simulated anneal, and if the result was still strongly discrepant, we excluded the source
from the analysis. This left us with 25 sources with similar PSF fit results to each other.

For each parameter in the PSF, we combined the best-fit values and uncertainties from these fits to produce
a probability distribution function which, due to the central limit theorem, we expect to be Gaussian in nature.
We then modelled this with a Gaussian function, to produce the best-fitting parameters used for 2SXPS, which were given in Table~\ref{tab:psf}.
Unlike the current CALDB parameters, we find that a Gaussian component is necessary; as can be seen
in Fig.~\ref{fig:psfmodel}, its inclusion improves the modelling of the PSF core, particularly at 8--15 pixels, which 
in turn allows the King component to broaden, better reproducing the wings. 

The PSF CALDB file allows for the PSF parameters to vary with energy, off-axis angle and the product of these properties.
Such variation, especially with energy, is expected physically, and was measured in the ground calibration \citep{Moretti04}.
All of the sources we analysed were close to on-axis; however, we split the data into different energy bands and repeated the fitting
process. Unfortunately, due to the lower number of counts per band the uncertainties on the parameters we derived were large. No evidence
was seen for energy dependence, but with no strong constraints. We elected therefore to treat the PSF as being independent of energy
for the purposes of our catalogue construction.
Since most modest-brightness sources are the target of their observation and are therefore on-axis, we lack the data
to properly calibrate the off-axis angle dependence, so this was also ignored.

% While the PSF for WT mode is not relevant to 2SXPS, sources in WT much are much brighter so in principle the PSF can be
% much better modelled. Calibrating the WT mode PSF revealed rather different parameters from the PC-mode PSF, most notably 
% the inclusion of the Gaussian component caused the PSF wings out beyond 100 pixel (\til3\arcmin) to be underestimated: in PC
% mode even stacking data the S/N was too low at this distance to be modelled.  Indeed {\bf Andy confirm please} the best-fitting model
% actually comprised two King components. The likely cause for this is that the true PSF is in fact not a simple King function, or King plus Gaussian,
% and thus the models we have fitted are approximations to the reality, and differ since the two modes tend to probe different
% areas of parameter space. In WT mode data, due both to the increased source
% brightness and the compression of the data onto a single axis, the PSF core (where the Gaussian offers such improvement in PC mode)
% is less important, whereas the distant wings (which cannot be seen in PC mode) play a role in constraining the PSF shape.
% This will be discussed in more detail in a CALDB release note accompanying new XRT PSF files.

\section{Catalog tables}
\label{sec:app_tabs}

There are four 2SXPS tables for download. The contents of these files are 
given in the following tables. The files are available in 3 formats: as a comma separated values (csv) file,
a FITS file, and as an SQL dump ({\sc mysql/mariadb} format).

The primary catalog product is the `2SXPS\_Sources' file which contains details of the unique sources, as described
in Table~\ref{tab:sourceDesc}. `2SXPS\_Datasets' (Table~\ref{tab:fieldInfoDesc}) describes the individual datasets; 
`2SXPS\_Detections' (Table~\ref{tab:detsDesc}) gives details of all of the individual detections and `2SXPS\_xCorr'
(Table~\ref{tab:xcorrDesc}) lists all the external catalog matches.

%%%%%%%%%%%%%%%%%%%%%%
%%% SOURCES TABLE %%%%
%%%%%%%%%%%%%%%%%%%%%%

{
\startlongtable
\begin{deluxetable*}{llll}
\tablecaption{Contents of the main catalog table (`sources'), containing an entry per unique source detected in 2SXPS.}
 \tablehead{
 \colhead{Field}     & \colhead{Units}   &       \colhead{Description} & \colhead{Has errors?$^1$} \\
}
\startdata
\multicolumn{4}{c}{\emph{Name and position}} \\
2SXPS\_ID               &          & Numerical unique source identifier within 2SXPS       &       \\ 
IAUName                 &          & IAU-format name, 2SXPS JHHMMSS[+-]ddmmsss       &       \\ 
RA                      &  Deg     & Right Ascension (J2000)            &       \\ 
Decl                    &  Deg     & Declination (J2000)                &       \\ 
Err90                   &  arcsec  & Position uncertainty, 90\% confidence, radial, assumed to be \\
                        &          & Rayleigh-distributed       &       \\ 
AstromType              &          & Provenance of source astrometry.\\
                        &          & 0=Swift star tracker, 1=XRT+2MASS \\
                        &          & astrometry       &       \\ 
l                       &  Deg     & Galactic longitude                 &       \\ 
b                       &  Deg     & Galactic latitude                  &       \\ 
MeanOffAxisAngle        &  arcmin  & The mean angular distance of the source from the XRT boresight \\
                        &          & in all observations in which the source was detected        \\ 
\bf OrigErr90           &  arcsec  & The (incorrect) error used for cross correlation \\
                        &          & (see footnote~\ref{fn:errerr}, p\pageref{fn:errerr}) \\
NearestNeighbour        & arcsec   & The distance to the closest 2SXPS source to this one       &       \\ 
NearestOKNeighbour      &  arcsec  & The distance to the closest 2SXPS source to this one which is \\
                        &          & ranked Good or Reasonable and has no other DetFlag bits set  &       \\
\multicolumn{4}{c}{\emph{Exposure details}} \\
Exposure                &  s       & The total exposure at the source location in the catalog       &       \\ 
FirstObsDate            &  UTC     & The time of the start of the first observation in 2SXPS which \\
                        &          & covered the source location       &       \\ 
LastObsDate             &  UTC     & The time of the end of the last observation in 2SXPS which \\
                        &          & covered the source location       &       \\ 
FirstObsMET             &  MET$^2$ & The time of the start of the first observation in 2SXPS which \\
                        &          & covered the source location       &       \\ 
LastObsMET              &  MET     & The time of the end of the last observation in 2SXPS which \\
                        &          & covered the source location       &       \\ 
FirstDetDate            &  UTC     & The date \& time of the start of the first observation in \\
                        &          & 2SXPS in which the source count rate is inconsisent with 0 at \\
                        &          & the 3-$\sigma$ level       &       \\ 
LastDetDate             &  UTC     & The date \& time of the end of the last observation in 2SXPS \\
                        &          & in which the source count rate is inconsisent with 0 at \\
                        &          & the 3-$\sigma$ level       &       \\ 
FirstDetMET             &  MET     & The time of the start of the first observation in 2SXPS in \\
                        &          & which the source count rate is inconsisent with 0 at the \\
                        &          & 3-$\sigma$ level     &       \\ 
LastDetMET              &  MET     & The time of the end of the last observation in 2SXPS in which \\
                        &          & the source count rate is inconsisent with 0 at the 3-$\sigma$ \\
                        &          & level      &       \\ 
FirstBlindDetDate       &  UTC     & The UTC date \& time of the start of the first observation in \\
                        &          & 2SXPS in which the source is detected in the blind search       &       \\ 
LastBlindDetDate        &  UTC     & The UTC date \& time of the end of the last observation in \\
                        &          & 2SXPS in which the source is detected in the blind search       &       \\ 
FirstBlindDetMET        &  MET     & The time of the start of the first observation in 2SXPS in  \\
                        &          & which the source is detected in the blind search,     &       \\ 
LastBlindDetMET         &  MET     & The time of the end of the last observation in 2SXPS in which \\
                        &          & the source is detected in the blind search.       &       \\ 
NumObs                  &          & The number of observations covering this source's position       &       \\ 
NumBlindDetObs          &          & The number of observations in which this source was found in \\
                        &          & a blind search.       &       \\ 
NumDetObs               &          & The number of observations in which this source is detected.       &       \\ 
BestDetectionID         &          & The ID of the detection from which the position and error \\
                        &          & were taken (cf the detections table).       &       \\ 
NonBlindDet\_band0      &  [Bool]  & Whether the count rate in the total band is inconsistent with \\
                        &          & 0 at the 3-$\sigma$ level (0 for no, 1 for yes).      &       \\ 
NonBlindDet\_band1      &  [Bool]  & Whether the count rate in the soft band is inconsistent with \\
                        &          & 0 at the 3-$\sigma$ level (0 for no, 1 for yes).       &       \\ 
NonBlindDet\_band2      &  [Bool]  & Whether the count rate in the medium band is inconsistent with \\
                        &          & 0 at the 3-$\sigma$ level (0 for no, 1 for yes).       &       \\ 
NonBlindDet\_band3      &  [Bool]  & Whether the count rate in the hard band is inconsistent with \\
                        &          & 0 at the 3-$\sigma$ level (0 for no, 1 for yes).       &       \\ 
\multicolumn{4}{c}{\emph{Flag details}} \\
DetFlag                 &          & The overall source detection flag       &       \\ 
FieldFlag               &          & The best field flag from all detections of this source       &       \\ 
DetFlag\_band0          &          & The overall detection flag the total band       &       \\ 
DetFlag\_band1          &          & The overall detection flag the soft band       &       \\ 
DetFlag\_band2          &          & The overall detection flag in the medium band       &       \\ 
DetFlag\_band3          &          & The overall detection flag in the hard band       &       \\ 
OpticalLoadingWarning   &  Mag     & The worst optical loading warning from all detections  \\
StrayLightWarning       & [Bool]   & Whether any detection of this source occurred within \\
                        &          & fitted stray light rings.       &       \\ 
NearBrightSourceWarning & [Bool]$^4$  & Whether any detection of this source occurred within \\
                        &          & the PSF wings of a bright object.       &       \\
IsPotentialAlias        &          & Whether the source is likely aliased with other sources& \\
PotentialAliasList      &          & The 2SXPS\_IDs of any sources which may be aliases of this&\\
\multicolumn{4}{c}{\emph{Count rate and variability information}} \\
Rate\_band0             & s$^{-1}$ & The mean count rate in the total band       &  Yes  \\ 
HR1                     &          & The aggregate HR1 hardness ratio of the source       &  Yes  \\ 
HR2                     &          & The aggregate HR2 hardness ratio of the source       &  Yes  \\ 
Rate\_band1             & s$^{-1}$ & The mean count rate in the soft band       &  Yes  \\ 
Rate\_band2             & s$^{-1}$ & The mean count rate in the medium band       &  Yes  \\ 
Rate\_band3             & s$^{-1}$ & The mean count rate in the hard band       &  Yes  \\ 
Counts\_band0           &          & The total number of counts in the source region in the total band       &       \\ 
Counts\_band1           &          & The total number of counts in the source region in the soft band       &       \\ 
Counts\_band2           &          & The total number of counts in the source region in the medium band       &       \\ 
Counts\_band3           &          & The total number of counts in the source region in the hard band       &       \\ 
BgCounts\_band0         &          & The total number of background counts expected in the source \\
                        &          & region in the total band       &       \\ 
BgCounts\_band1         &          & The total number of background counts expected in the source \\
                        &          & region in the soft band       &       \\ 
BgCounts\_band2         &          & The total number of background counts expected in the source \\
                        &          & region in the medium band       &       \\ 
BgCounts\_band3         &          & The total number of background counts expected in the source \\
                        &          & region in the hard band       &       \\ 
RateCF\_band0           &          & The PSF correction factor in the total band       &       \\ 
RateCF\_band1           &          & The PSF correction factor in the soft band       &       \\ 
RateCF\_band2           &          & The PSF correction factor in the medium band       &       \\ 
RateCF\_band3           &          & The PSF correction factor in the hard band       &       \\ 
UL\_band0               & s$^{-1}$ & The 3-$\sigma$ upper limit on the count rate in the total band       &       \\ 
UL\_band1               & s$^{-1}$ & The 3-$\sigma$ upper limit on the count rate in the soft band       &       \\ 
UL\_band2               & s$^{-1}$ & The 3-$\sigma$ upper limit on the count rate in the medium band       &       \\ 
UL\_band3               & s$^{-1}$ & The 3-$\sigma$ upper limit on the count rate in the hard band       &       \\ 
PeakRate\_band0$^5$     & s$^{-1}$ & The peak count rate in the total band       &  Yes  \\ 
PeakRate\_band1$^5$     & s$^{-1}$ & The peak count rate in the soft band       &  Yes  \\ 
PeakRate\_band2$^5$     & s$^{-1}$ & The peak count rate in the medium band       &  Yes  \\ 
PeakRate\_band3$^5$     & s$^{-1}$ & The peak count rate in the hard band       &  Yes  \\ 
PvarPchiSnapshot\_band0 &          & The probability that the source count rate in the total band  \\
                        &          & does not vary between snapshots       &       \\ 
PvarPchiSnapshot\_band1 &          & The probability that the source count rate in the soft band  \\
                        &          & does not vary between snapshots       &       \\ 
PvarPchiSnapshot\_band2 &          & The probability that the source count rate in the medium band \\
                        &          & does not vary between snapshots       &       \\ 
PvarPchiSnapshot\_band3 &          & The probability that the source count rate in the hard band \\
                        &          & does not vary between snapshots       &       \\ 
PvarPchiSnapshot\_HR1   &          & The probability that the source HR1 hardness ratio does not \\
                        &          & vary between snapshots       &       \\ 
PvarPchiSnapshot\_HR2   &          & The probability that the source HR2 hardness ratio does no \\
                        &          &t vary between snapshots       &       \\ 
PvarPchiObsID\_band0    &          & The probability that the source count rate in the total band \\
                        &          & does not vary between observations       &       \\ 
PvarPchiObsID\_band1    &          & The probability that the source count rate in the soft band \\
                        &          & does not vary between observations       &       \\ 
PvarPchiObsID\_band2    &          & The probability that the source count rate in the medium band \\
                        &          & does not vary between observations       &       \\ 
PvarPchiObsID\_band3    &          & The probability that the source count rate in the hard band \\
                        &          & does not vary between observations       &       \\ 
PvarPchiObsID\_HR1      &          & The probability that the source HR1 hardness ratio does not \\
                        &          & vary between observations       &       \\ 
PvarPchiObsID\_HR2      &          & The probability that the source HR2 hardness ratio does not \\
                        &          & vary between observations       &       \\ 
\multicolumn{4}{c}{\emph{Flux and spectral information}} \\
GalacticNH              &   \cms   & The Galactic absorption column in the direction of the source, \\
                        &          & from Willingale et al (2013) \\
WhichPow                &          & Which method of determining the spectral properties assuming \\
                        &          & a power-law was used       &       \\ 
WhichAPEC               &          & Which method of determining the spectral properties assuming \\
                        &          & an APEC was used   &       \\ 
PowECFO                 & \ecf     & The observed flux ECF$^3$, assuming a power-law spectrum.       &       \\ 
PowECFU                 & \ecf     & The unabsorbed flux ECF, assuming a power-law spectrum.       &       \\ 
PowFlux                 & \flux    & The mean total observed flux assuming a power-law spectrum.       &  Yes  \\ 
PowUnabsFlux            & \flux    & The mean total unabsorbed flux assuming a power-law spectrum.       &  Yes  \\ 
APECECFO                & \ecf     & The observed flux ECF, assuming an APEC spectrum.       &       \\ 
APECECFU                &  \ecf    & The unabsorbed flux ECF, assuming an APEC spectrum.       &       \\ 
APECFlux                & \flux    & The mean total observed flux assuming an APEC spectrum.       &  Yes  \\ 
APECUnabsFlux           & \flux    & The mean total unabsorbed flux assuming an APEC spectrum.       &  Yes  \\ 
PowPeakFlux             & \flux    & The peak total observed flux assuming a power-law spectrum.       &  Yes     \\ 
PowPeakUnabsFlux        & \flux    & The peak total unabsorbed flux assuming a power-law spectrum.       &  Yes     \\ 
APECPeakFlux            & \flux    & The peak total observed flux assuming an APEC spectrum.       &  Yes     \\ 
APECPeakUnabsFlux       & \flux    & The peak total unabsorbed flux assuming an APEC spectrum.       &  Yes     \\ 
FixedPowECFO            & \ecf     & The observed flux ECF, assuming the canned power-law spectrum.       &       \\ 
FixedPowECFU            & \ecf     & The unabsorbed flux ECF, assuming the canned power-law spectrum.       &       \\ 
FixedPowFlux            & \flux    & The mean total observed flux assuming the canned power-law spectrum.       &  Yes     \\ 
FixedPowUnabsFlux       & \flux    & The mean total unabsorbed flux assuming the canned power-law spectrum.       &  Yes     \\ 
FixedAPECECFO           & \ecf     & The observed flux ECF, assuming the canned APEC spectrum.       &       \\ 
FixedAPECECFU           & \ecf     & The unabsorbed flux ECF, assuming the canned APEC spectrum.       &       \\ 
FixedAPECFlux           & \flux    & The mean total observed flux assuming the canned APEC spectrum.       &  Yes     \\ 
FixedAPECUnabsFlux      & \flux    & The mean total unabsorbed flux assuming the canned APEC spectrum.       &  Yes     \\ 
InterpPowECFO           & \ecf     & The observed flux ECF, assuming the power-law spectrum \\
                        &          & interpolated from the HRs.       &       \\ 
InterpPowECFU           & \ecf     & The unabsorbed flux ECF, assuming the power-law spectrum \\
                        &          & interpolated from the HRs.       &       \\ 
InterpPowNH             & \cms     & The hydrogen column density inferred assuming the power-law \\
                        &          & spectrum interpolated from the HRs.       &  Yes  \\ 
InterpPowGamma          &          & The spectral photon index inferred assuming the power-law \\
                        &          & spectrum interpolated from the HRs.       &  Yes  \\ 
InterpPowFlux           &  \flux   & The mean total observed flux assuming the power-law spectrum  \\
                        &          &interpolated from the HRs.       &  Yes \\ 
InterpPowUnabsFlux      &  \flux   & The mean total unabsorbed flux assuming the power-law spectrum \\
                        &          & interpolated from the HRs.       & Yes   \\ 
InterpAPECECFO          &  \ecf    & The observed flux ECF, assuming the APEC spectrum interpolated \\
                        &          & from the HRs.       &       \\ 
InterpAPECECFU          &  \ecf    & The unabsorbed flux ECF, assuming the APEC spectrum interpolated \\
                        &          & from the HRs.       &       \\ 
InterpAPECNH            &  \cms    & The hydrogen column density inferred assuming the APEC spectrum \\
                        &          & interpolated from the HRs.       &  Yes  \\ 
InterpAPECkT            &  keV     & The temperature inferred assuming the APEC spectrum interpolated \\
                        &          & from the HRs.       &  Yes  \\ 
InterpAPECFlux          &  \flux   & The mean total observed flux assuming the APEC spectrum \\
                        &          & interpolated from the HRs.       &  Yes  \\ 
InterpAPECUnabsFlux     &  \flux   & The mean total unabsorbed flux assuming the APEC spectrum \\
                        &          & interpolated from the HRs.       & Yes   \\ 
P\_pow                  &          &  The probability that the HR values of this source could be \\
                        &          & obtained if the true spectrum is an absorbed power-law       &       \\ 
P\_APEC                 &          &  The probability that the HR values of this source could be \\
                        &          & obtained if the true spectrum is an APEC.       &       \\ 
FittedPowECFO           &  \ecf    & The observed flux ECF, assuming the power-law spectral model \\
                        &          & fitted to a custom-built spectrum.       &       \\ 
FittedPowECFU           &  \ecf    & The unabsorbed flux ECF, assuming the power-law spectral model \\
                        &          & fitted to a custom-built spectrum.       &       \\ 
FittedPowNH             &  \cms    & The hydrogen column density inferred assuming the power-law \\
                        &          & spectral model fitted to a custom-built spectrum.       & Yes   \\ 
FittedPowGamma          &          & The spectral photon index inferred assuming the power-law \\
                        &          & spectral model fitted to a custom-built spectrum.       & Yes   \\ 
FittedPowFlux           &  \flux   & The mean total observed flux assuming the power-law spectral \\
                        &          & model fitted to a custom-built spectrum.       & Yes   \\ 
FittedPowUnabsFlux      & \flux    & The mean total unabsorbed flux assuming the power-law spectral \\
                        &          & model fitted to a custom-built spectrum.       & Yes   \\ 
FittedPowCstat          &          & The C-statistic from the power-law spectral fit to the \\
                        &          & custom-built spectrum.       &       \\ 
FittedPowDOF            &          & The number of degrees of freedom in the power-law spectral \\
                        &          & fit to the custom-built spectrum.       &       \\ 
FittedPowReducedChi2    &          & The Churazov-weighted reduced \chisq\ from the power-law \\
                        &          & spectral fit to the custom-built spectrum.       &       \\ 
FittedAPECECFO          &   \ecf   & The observed flux ECF, assuming the APEC spectral model \\
                        &          & fitted to a custom-built spectrum.       &       \\ 
FittedAPECECFU          &  \ecf    & The unabsorbed flux ECF, assuming the APEC spectral model \\
                        &          & fitted to a custom-built spectrum.       &       \\ 
FittedAPECNH            &  \cms    & The hydrogen column density inferred assuming the APEC \\
                        &          & spectral model fitted to a custom-built spectrum.       & Yes   \\ 
FittedAPECkT            &  keV     & The temperature inferred assuming the APEC spectral model \\
                        &          & fitted to a custom-built spectrum.       & Yes   \\ 
FittedAPECFlux          &  \flux   & The mean total observed flux assuming the APEC spectral \\
                        &          & model fitted to a custom-built spectrum.       & Yes   \\ 
FittedAPECUnabsFlux     &  \flux   & The mean total unabsorbed flux assuming the APEC spectral \\
                        &          & model fitted to a custom-built spectrum.       &  Yes  \\ 
FittedAPECCstat         &          & The C-statistic from the APEC spectral fit to the \\
                        &          & custom-built spectrum.       &       \\ 
FittedAPECDOF           &          & The number of degrees of freedom in the APEC spectral fit \\
                        &          & to the custom-built spectrum.       &       \\ 
FittedAPECReducedChi2   &          & The Churazov-weighted reduced \chisq\ from the APEC spectral \\
                        &          & fit to the custom-built spectrum.       &       \\ 
HasSpec                 &          & Whether a custom-built spectrum was created for this source.       &       \\ 
\multicolumn{4}{l}{\emph{Cross-correlation information}} \\
NumExternalMatches      &          & The number of external sources found to agree spatially with \\
                        &          & this one at the 3-$\sigma$ level.       &       \\ 
NumExternalMatches\_slim&          & The number of external sources found to agree spatially with \\
                        &          & this one at the 3-$\sigma$ level, excluding 2MASS, USNO-B1 \\
                        &          & and ALLWISE matches.       &       \\ 
MatchInROSHRI           &  [Bool]  & Whether the source has a match in ROSAT HRI        &       \\ 
MatchIn2RXS             &  [Bool]  & Whether the source has a match in 2RXS        &       \\ 
MatchIn3XMMDR8          &  [Bool]  & Whether the source has a match in 3XMM-DR8        &       \\ 
MatchIn3XMM\_Stack      &  [Bool]  & Whether the source has a match in 3XMM-DR7s       &       \\ 
MatchInXMMSL2           &  [Bool]  & Whether the source has a match in XMMSL2        &       \\ 
MatchInSwiftFT          &  [Bool]  & Whether the source has a match in SwiftFT        &       \\ 
MatchIn1SWXRT           &  [Bool]  & Whether the source has a match in 1SWXRT        &       \\ 
MatchInXRTGRB           &  [Bool]  & Whether the source has a match in the XRT GRB afterglows.       &       \\ 
MatchInSDSSQSO          &  [Bool]  & Whether the source has a match in SDSS QSO DR14        &       \\ 
MatchIn2MASS            &  [Bool]  & Whether the source has a match in 2MASS        &       \\ 
MatchInUSNOB1           &  [Bool]  & Whether the source has a match in USNO-B1        &       \\ 
MatchIn2CSC             &  [Bool]  & Whether the source has a match in 2CSC        &       \\ 
MatchIn1SXPS            &  [Bool]  & Whether the source has a match in 1SXPS        &       \\ 
MatchInALLWISE          &  [Bool]  & Whether the source has a match in ALLWISE        &       \\ 
\enddata                   
\label{tab:sourceDesc}
\tablecomments{
Boolean columns (marked as `[Bool]' above) have a value of 0 for false and 1 for true. \\
$^1$ This is `no' unless stated. For a field with errors, there are two error fields, \emph{fieldname}\_pos and 
\emph{fieldname}\_neg.  \\
$^2$ MET = Swift Mission Elapsed Time = Seconds since 2001 Jan 01 00:00:00 (TT) \\
$^3$ ECF = Energy Conversion Factor, i.e.\ the conversion from observed 0.3--10 keV counts to 0.3--10 keV flux; ECFs
are provided to convert to observed and unabsorbed flux. \\
$^4$ NearBrightSourceWarning can have a value of 2, as discussed in Section~\ref{sec:Ltest}. \\
$^5$ The peak rate was defined in Section~\ref{sec:prods}.
}
\end{deluxetable*}
}

%%%%%%%%%%%%%%%%%%%%%%%
%%% DATASETS TABLE %%%%
%%%%%%%%%%%%%%%%%%%%%%%

\startlongtable
\begin{deluxetable*}{lll}
\label{tab:fieldInfoDesc}
\tablecaption{Contents of the `Datasets' catalog table, containing an entry per dataset in the catalog}
 \tablehead{
 \colhead{Field}     & \colhead{Units}   &       \colhead{Description}  \\
}
\startdata
ObsID$^1$         &                           & \swift\ obsID of the dataset   \\
FieldFlag         &                           & The warning flag associated with this dataset   \\
RA                &  deg                      & The Right Ascension (J2000) of the dataset center   \\
Decl              &  deg                      & The declination (J2000) of the dataset center\\
l                 &  deg                      & Galactic longitude of the dataset center   \\
b                 &  deg                      & Galactic latitude of the dataset center   \\
ImageSize         &  pix                      & The side length of the dataset image in XRT pixels   \\
ExposureUsed      &   s                       & The post-filtering exposure in the dataset   \\
OriginalExposure  &   s                       & The original exposure in the dataset   \\
StartTime\_MET     &  MET                      & The start time of the dataset   \\
StopTime\_MET      &  MET                      & The end time of the dataset   \\
MidTime\_MET       &  MET                      & The mid-time of the dataset   \\
MidTime\_TDB       &  TDB                      & The mid-time of the dataset   \\
MidTime\_MJD       &  MJD                      & The mid-time of the dataset   \\
StartTime\_UTC     &  UTC                      & The start time of the dataset   \\
StopTime\_UTC      &  UTC                      & The end time of the dataset   \\
FieldBG\_band0     &  ct s$^{-1}$ pix$^{-1}$   & The mean background level in the total band   \\
FieldBG\_band1     &  ct s$^{-1}$ pix$^{-1}$   & The mean background level in the soft band   \\
FieldBG\_band2     &  ct s$^{-1}$ pix$^{-1}$   & The mean background level in the medium band   \\
FieldBG\_band3     &  ct s$^{-1}$ pix$^{-1}$   & The mean background level in the hard band   \\
NumSrc\_band0      &                           & The number of sources detected in this dataset   \\
                  &                            & in the total band   \\
NumOK\_band0       &                           & The number of \emph{Good} or \emph{Reasonable} \\
                  &                            & sources detected in this dataset in the total band   \\
MedianDist\_band0  &  arcsec                   & The median distance between sources detected in   \\
                  &                            & this dataset in the total band   \\
NumSrc\_band1      &                           & The number of sources detected in this dataset in \\
                  &                            & the soft band   \\
NumOK\_band1       &                           & The number of good or reasonable sources detected \\
                  &                            & in this dataset in the soft band   \\
MedianDist\_band1  &  arcsec                   & The median distance between sources detected in   \\
                  &                            &  this dataset in the soft band   \\
NumSrc\_band2      &                           & The number of sources detected in this dataset   \\
                  &                            &  in the medium band   \\
NumOK\_band2       &                           & The number of good or reasonable sources detected   \\
                  &                            &  in this dataset in the medium band   \\
MedianDist\_band2  &  arcsec                   & The median distance between sources detected in   \\
                   &                           &  this dataset in the medium band   \\
NumSrc\_band3      &                           & The number of sources detected in this dataset in   \\
                   &                           &  the hard band   \\
NumOK\_band3       &                           & The number of good or reasonable sources detected   \\
                  &                            &  in this dataset in the hard band   \\
MedianDist\_band3  &  arcsec                   & The median distance between sources detected in   \\
                  &                            &  this dataset in the hard band   \\
NumberOfSnapshots &                            & The number of snapshots contributing to this dataset   \\
AstromError       &  arcsec                    & The 90\% confidence radial uncertainty on the   \\
                  &                            & XRT-2MASS astrometric solution   \\
CRVAL1\_corr       &                           & The CRVAL1 WCS reference value for the dataset   \\
                  &                            & derived from the XRT-2MASS astrometric solution   \\
CRVAL2\_corr       &                           & The CRVAL2 WCS reference value for the dataset   \\
                  &                            & derived from the XRT-2MASS astrometric solution   \\    
CROTA2\_corr       &                           & The CROTA1 WCS reference value for the dataset   \\
                  &                            & derived from the XRT-2MASS astrometric solution   \\
\enddata
\tablecomments{$^1$ Values $>10^{10}$ refer to stacked images.}
\end{deluxetable*}

%%%%%%%%%%%%%%%%%%%%%%%%%
%%% DETECTIONS TABLE %%%%
%%%%%%%%%%%%%%%%%%%%%%%%%

\startlongtable
\begin{deluxetable*}{llll}
\tablecaption{Contents of the `Detections' catalog table, containing an entry per detection in the catalogue}
 \tablehead{
 \colhead{Field}     & \colhead{Units}   &       \colhead{Description} & \colhead{Has errors?} \\
}
\startdata
DetectionID             &           & A unique identifier for this detection     \\
2SXPS\_ID               &           & The 2SXPS sourceID with which this detection is associated     \\
SourceNo                &           & The identifier of this source within this obsid and band     \\
Band                    &           & The energy band in which this detection occurred   \\
ObsID                   &           & The identifier of the observation or stacked image in which \\
                        &           & this detection occurred.     \\
CorrectedExposure       &  s        & The exposure time at the position of the source in this obsID \\
ExposureFraction        &  s        & The fractional exposure at the position of this source, \\
                        &           & i.e. the exposure divided by the nominal exposure for the field     \\
OffaxisAngle            &  arcmin   & The angular distance of the source from the XRT boresight \\
RA                      &  deg      & Right Ascension (J2000)   & Yes     \\
Decl                    &  deg      & Decliniation (J2000)   & Yes     \\
Err90                   &  arcsec   & Position uncertainty, 90\% confidence, radius\\
RA\_corrected           &  deg      & Right Ascension (J2000) using XRT-2MASS astrometry     \\
Decl\_corrected         &  deg      & Declination (J2000)  using XRT-2MASS astrometry     \\
Err90\_corrected        &  arcsec   & Uncertainty on the  position, 90\% confidence radius \\
l                       &  deg      & Galactic longitude    \\
b                       &  deg      & Galactic latitude     \\
l\_corrected            &  deg      & Galactic longitude  using XRT-2MASS astrometry     \\
b\_corrected            &  deg      & Galactic latitude  using XRT-2MASS astrometry     \\
IMG\_X                  &           & The x position of the object in the SKY image plane     \\
IMG\_Y                  &           & The y position of the object in the SKY image plane     \\
NearestNeighbour        &  arcsec   & The distance to the closest detection to this one, in this image.     \\
NearestOKNeighbour      &  arcsec   & The distance to the closest Good or Reasonable detection to \\
                        &           & this one, in this image.     \\
DetFlag                 &           & The detection flag    \\
OpticalLoadingWarning   &  mag      & Optical loading warning level   \\
StrayLightWarning       &           & Whether this  detection occurred within fitted stray light rings.     \\
NearBrightSourceWarning &           & Whether this detection  occurred within the PSF wings of a \\
                        &           & fitted bright source     \\
MatchesKnownExtended    &           & Whether the position of this source matches a known \\
                  &                            & extended X-ray source.     \\
PileupFitted            &           & Whether the accepted fit included pile up.     \\
SNR                     &           & The signal to noise ratio of the detection.     \\
CtsInPSFFit             &           & Number of counts in the image region over which the final \\
                        &           & PSF fit was performed     \\
BGRateInPSFFit          &           & Mean count rate in the background map in the region over \\
                        &           & which the final PSF fit was performed     \\
Cstat                   &           & \cstat\ from the PSF fit     \\
Cstat\_nosrc            &           & \cstat\ value if no source is fitted     \\
L\_src                  &           & The likelihood value that this detection is not just \\
                        &           & a background fluctuation.     \\
Cstat\_flat             &           & \cstat\ assuming a spatially uniform increase above the background \\
Lflat                   &           & The likelihood value that this detection is PSF like, not flat  \\
FracPix                 &           & The fraction of pixels within the PSF fit region \\
                        &           & which are exposed.     \\
Pileup\_S               &           & The best-fitting S parameter of the pile up model.     \\
Pileup\_l               &           & The best-fitting l parameter of the pile up model.     \\
Pileup\_c               &           & The best-fitting c parameter of the pile up model.     \\
Pileup\_tau             &           & The best-fitting tau parameter of the pile up model.     \\
Cstat\_altPileup        &           & \cstat\ from the unusued fit. i.e. if the piled up model was used, \\
                        &           & this gives the Cstat from the non-piled-up fit, and vice-versa.     \\
PSF\_Fit\_Radius        &  pix      & The radius of the circular region over which PSF fitting \\
                        &           & was carried out    \\   
CellDetect\_BoxWidth    &  pix      & The full width of the cell-detect box in which this \\
                        &           & source was detected    \\   
Rate                    &  s$^{-1}$ & The count rate of this detection    & Yes     \\
CtsInRate               &           & The total number of counts in the region used to \\
                        &           & extract the count rate     \\
BGCtsInRate             &           & The total number of counts in the region used to \\
                        &           & extract the count rate     \\
Rate\_CF                &           & The PSF correction factor for the count rate     \\
BGRateInRate            &  s$^{-1}$ & The background rate in the region used for count rate \\
                        &           & calculation.     \\
OrigErr90               &  arcsec   & The (incorrect) error used for making the unique source list\\
                        &           & (see footnote~\ref{fn:errerr}, p\pageref{fn:errerr}) \\
OrigErr90\_Corrected    &  arcsec   & The (incorrect) astrometrically-corrected error used for making \\
                        &          &  the unique source list (see footnote~\ref{fn:errerr}, p\pageref{fn:errerr}) \\
\enddata
\label{tab:detsDesc}
\end{deluxetable*}

%%%%%%%%%%%%%%%%%%%%
%%% XCORR TABLE %%%%
%%%%%%%%%%%%%%%%%%%%

\begin{deluxetable*}{lll}
\label{tab:xcorrDesc}
\tablecaption{Contents of the `Cross Correlations' catalog table, containing an entry for every match between
a 2SXPS source and a source from another catalog}
 \tablehead{
 \colhead{Field}     & \colhead{Units}   &       \colhead{Description}  \\
}
\startdata   
2SXPS\_ID  &         & The 2SXPS sourceID \\
ExtCat\_ID &         & The name of the source in the external catalog \\
Catalogue  &         & The catalog containing the matched source \\
Distance   & arcsec  & The distance between the 1SXPS source and \\
           &         & external catalog source \\
RA         & degrees & The RA (J2000) of the source in the external catalog \\
Decl       & degrees & The Declination (J2000) of the source in the external catalog \\
Err90      & arcsec  & The 90\%\ confidence radial uncertainty in the external \\ 
           &         & catalog position (inc systematics) \\
\enddata
\end{deluxetable*}

\label{lastpage}

\end{document}